\documentclass[aps,pre,twocolumn,superscriptaddress]{revtex4-1}

\usepackage{graphicx}
\usepackage{epstopdf}
\usepackage{ucs}

\usepackage{amssymb}
\usepackage{epstopdf}
\usepackage{amsmath}
\usepackage{amsthm}
\usepackage{color}

\newcount\colveccount
\newcommand*\colvec[1]{
        \global\colveccount#1
        \begin{pmatrix}
        \colvecnext
}
\def\colvecnext#1{
        #1
        \global\advance\colveccount-1
        \ifnum\colveccount>0
                \\
                \expandafter\colvecnext
        \else
                \end{pmatrix}
        \fi
}

\def\be{\begin{equation}}   \def\ee{\end{equation}}

\usepackage{bm}

\begin{document}
%\significancestatement{Many blood cells have a discoidal shape, which is essential for them to function properly within the organism. For some blood cells, such as blood platelets, this shape is due to the interplay between the elasticity of the marginal band, which is a closed ring of stiff filaments called microtubules, and the tension of the cell cortex, a polymer scaffold associated with the plasma membrane. Dmitrieff et al. explain how cell shape is determined by the mechanical balance between these two components. Remarkably, this theory is confirmed over nearly three orders of magnitude, by data collected from 25 species. The theory also shows how the composite structure is adapted to resist transient mechanical challenge, such as the shear encountered in the blood stream.}

\title{Balance of microtubule stiffness and cortical tension determines the size of blood cells with marginal band across species}
\author{Serge Dmitrieff} \author{ Adolfo Alsina} \author{ Aastha Mathur} \author{ Fran\c{c}ois N\'{e}d\'{e}lec}
\affiliation{Cell Biology and Biophysics Unit, European Molecular Biology Laboratory, Meyerhofstrasse 1, 69117 Heidelberg, Germany.}
%\author[a]{Serge Dmitrieff}
%\author[a]{Adolfo Alsina}
%\author[a]{Aastha Mathur}
%\author[a,1]{François Nédélec}
%\affil[a]{Cell Biology and Biophysics Unit, European Molecular Biology Laboratory, Meyerhofstrasse 1, 69117 Heidelberg, Germany.}

%\authorcontributions{Conceived and designed the work : FN SD. Performed the theory : SD AA. Performed the experiments : AM. Analyzed data : SD AA AM. Wrote the paper : FN SD AM.}
%\authordeclaration{The authors declare no conflict of interest.}
%\correspondingauthor{\textsuperscript{1}To whom correspondence should be addressed. E-mail: nedelec@embl.de}
%\keywords{Platelets $|$ Red blood cells $|$ Mechanics $|$ Microtubules $|$ Actin $|$ Theory $|$ Scaling} 

\date{This manuscript was compiled on \today}
%\doi{\url{www.pnas.org/cgi/doi/10.1073/pnas.XXXXXXXXXX}}

\begin{abstract}
The fast blood stream of animals is associated with large shear stresses. Consequently, blood cells have evolved a special morphology and a specific internal architecture allowing them to maintain their integrity over several weeks.
For instance, non-mammalian red blood cells, mammalian erythroblasts and platelets have a peripheral ring of microtubules, called the marginal band, that flattens the overall cell morphology by pushing on the cell cortex. 
In this article, we model how the shape of these cells stems from the balance between marginal band elasticity and cortical tension. 
We predict that the diameter of the cell scales with the total microtubule polymer, and verify the predicted law across a wide range of species. 
Our analysis also shows that the combination of the marginal band rigidity and cortical tension increases the ability of the cell to withstand forces without deformation. 
Finally, we model the marginal band coiling that occurs during the disc-to-sphere transition observed for instance at the onset of blood platelet activation.
We show that when cortical tension increases faster than crosslinkers can unbind, the marginal band will coil, whereas if the tension increases slower, the marginal band may shorten as microtubules slide relative to each other.
\end{abstract}

%\maketitle

%\verticaladjustment{-2pt}

\maketitle
%\thispagestyle{firststyle}
%\ifthenelse{\boolean{shortarticle}}{\ifthenelse{\boolean{singlecolumn}}{\abscontentformatted}{\abscontent}}{}

%\section*{Introduction}

\begin{figure*}[t]
\hspace{1cm}\includegraphics[width=0.9\textwidth]{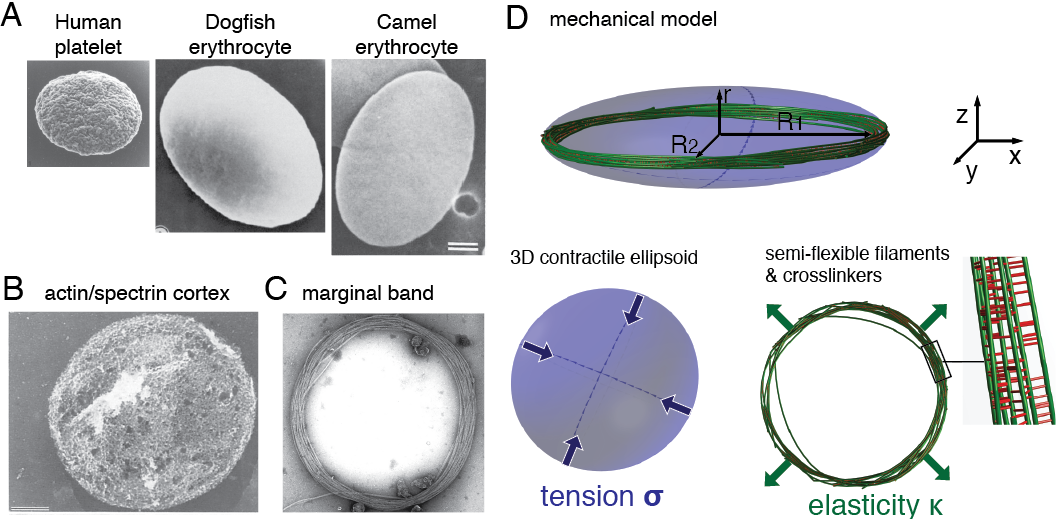}  \\
\caption{
\label{illus_intro}
A) Scanning electron micrographs of Platelets and Erythrocytes shown at the same scale \cite{white2013,joseph1984cytoskeletal,schroter1990influence}, scale bar $1\mu m$. 
B) The actin/spectrin cortex of platelets, EM from \cite{hartwig1991cytoskeleton}, scale bar $0.5 \mu m$. 
C) The MB of platelets is made of multiple MTs bundled by motors and crosslinkers \cite{bathe2008cytoskeletal},  EM from \cite{white2013}. 
D) In our model, the shape of the cell is determined by the balance of two forces. Because of microtubule stiffness $\kappa$, the MB pushes against the tense cortex, which resists by virtue of its surface tension $\sigma$. 
}
\end{figure*}

The shape of animal cells is determined by the cytoskeleton, including microtubules (MTs), contractile networks of actin filaments, intermediate filaments and other mechanical elements. The 3D geometry of most cells in a multi-cellular organism is also largely determined by their adhesion to neighbouring cells or to the extra-cellular matrix \cite{lecuit2007cell}. This is not however the case for blood cells as they circulate freely within the fluid environment of the blood plasma.
Red blood cells (RBC) and thrombocytes in non-mammalian animals \cite{goniakowska1976evidence,lee2004shape}, platelets and erythroblasts in mammals \cite{patel2008visualization,van1973microtubule} adopt a simple ellipsoidal shape (Fig. \ref{illus_intro}A). 
This shape is determined by two components: a ring of MTs, called the marginal band (MB), and a protein cortex at the cell periphery.

In the case of platelets and non-mammalian red blood cells, both components are relatively well characterized (Fig. \ref{illus_intro}). The cortex is a composite structure made of spectrin, actin and intermediate filaments (Fig. \ref{illus_intro}B), and its complex architecture is likely to be dynamic \cite{patel2011spectrin,thon2012microtubule,cohen1982cytoskeletal}. 
It is a thin network under tension \cite{evans1989apparent}, that on its own would lead to a spherical morphology \cite{stewart2011hydrostatic}.
This effect is counterbalanced by the MB, a ring made of multiple dynamic MTs, held together by crosslinkers and molecular motors into a closed circular bundle \cite{patel2008visualization, bender2015microtubule} (Fig. \ref{illus_intro}C).
The MB is essential to maintain the flat morphology, and treatment with a MT destabilising agent causes platelets to round up \cite{white1998microtubule}.
It was also reported that when the cell is activated, the MB is often seen to buckle \cite{lee2004shape}. This phenomenon is reminiscent of the buckling of a closed elastic ring \cite{ostermeir2010buckling}, but the MB is not a continuous structure of constant length.

Indeed, an important feature of the MB is that is it made of multiple MTs, connected by dynamic crosslinkers. The rearrangement of connectors could allow MTs to slide relative to one another, and thus would allow the length of the MB to change. Secondly, MT growth or depolymerisation would also induce reorganisation. However, in the absence of sliding, elongation or shortening of single MTs would principally affect the thickness of the MB ({\it i.e.} the number of MT in the cross-section) rather than its length. It was also suggested that molecular motors may drive the elongation of the MB \cite{diagouraga2014motor}, but this possibility remains mechanistically unclear.
These aspects have received little attention so far, and much remains to be done before we can understand how the original architecture of these cells is  adapted to their unusual environment, and to the mechanical constraints associated with it \cite{joseph1984cytoskeletal}.

%Sliding of the MTs relative to each other can possibly cause the MB to change length, whereas MT growth or depolymerisation would principally affect its thickness. 
%Thus subjected to the constraints of the cortex, the MB could either shorten, if the MT slide relative to each other, or buckle if the forces exceed a certain threshold. 
%For the MTs to slide relative to each other, the connectors linking them into a circular bundle would have to rearrange, and thus the propensity to shorten is indirectly determined by the molecular properties of the connectors.
%It is also possible for the MB to elongate \cite{diagouraga2014motor}, for example if the molecular motors that are bridging the MTs shear them appropriately to do so, but how this might occur remains unclear.

We argue here that despite the potential complexity of the system, the equilibrium between MB elasticity and cortical tension can be understood in simple mechanical terms. 
We first predict that the main cell radius should scale with the total length of MT polymer and inversely with the cortical tension, and test the predicted relationship using data from a wide range of species. 
We then simulate the shape changes observed during platelets activation \cite{kuwahara2002platelet}, 
discussing that a rapid increase of tension leads to MB coiling accompanied by a shortening of the ring, while a slow increase of tension leads to a shortening of the ring without coiling.
Finally, by computing the buckling force of a ring confined within an ellipsoid, we find that the resistance of the cell to external forces is dramatically increased compared to the resistance of the ring alone.

\section*{Results}

\subsection*{Cell size is controlled by total microtubule polymer and cortical tension}

\begin{figure}[t]
\includegraphics[width=8.7cm]{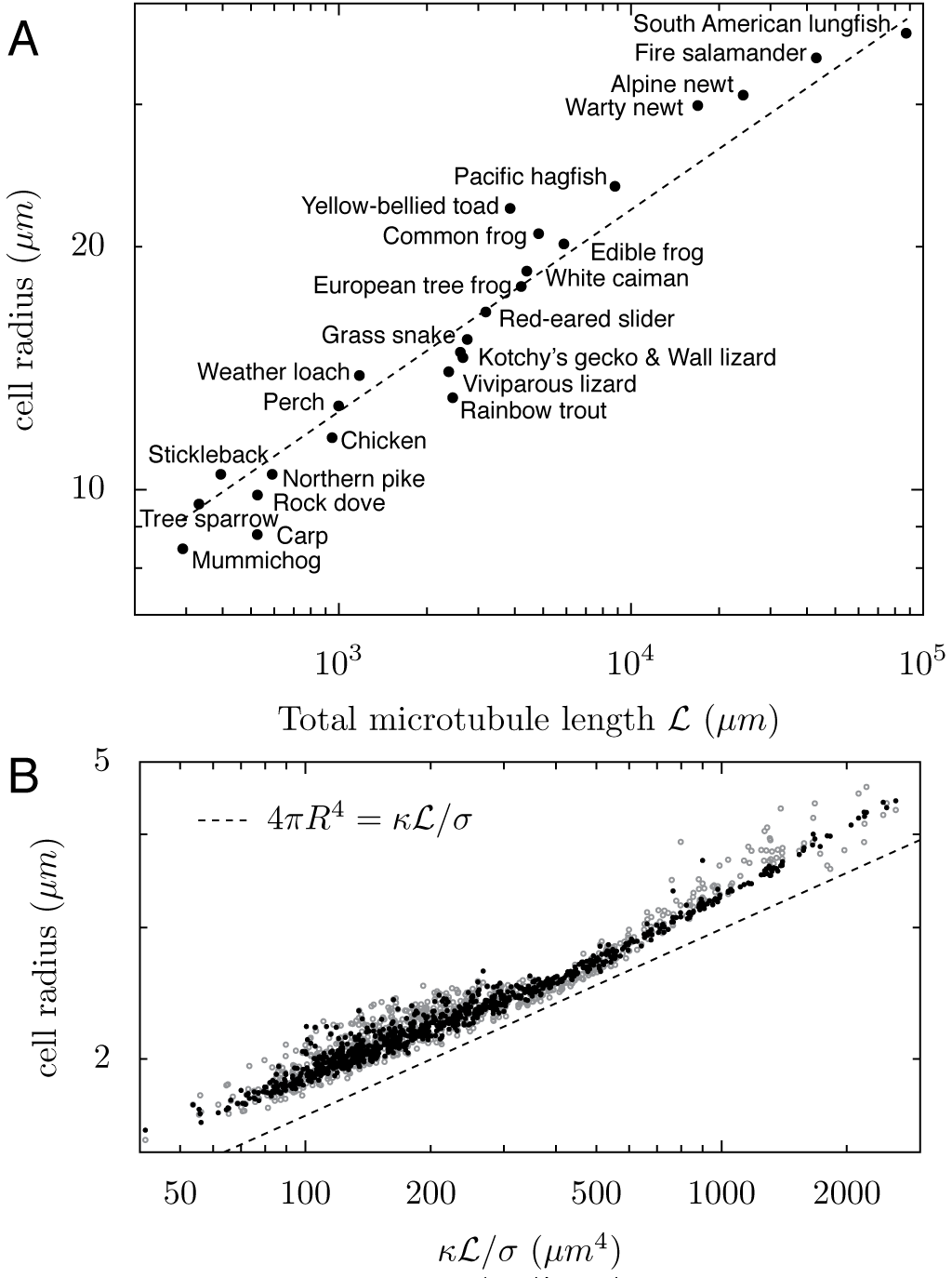}  \\
\caption{
\label{MB_radius}
A) Cell radius as a function of total MT length $\mathcal{L}$.
Dots: data from 25 species (\cite{goniakowska1976evidence}). $\mathcal{L} $ was estimated from the number of microtubules in a cross-section, measured in electron microscopy, and the cell radius. 
B) Cell radius as a function of $\mathcal{L} \kappa / \sigma$ in simulations with $0$ (gray points) or $10000$ (black points) crosslinkers. On both graphs, the dashed line indicates the theory $4 \pi R^4 =  \kappa \mathcal{L} / \sigma$.
}
\end{figure}

We first apply scaling arguments to explore how cell shape is determined by the mechanical equilibrium between MB elasticity and cortical tension. 
%Specifically, the time scale we consider exceeds connectors binding and unbinding dynamics, allowing us to ignore the mechanical contribution of the connectors such as crosslinkers.
%In this part, we ignore MT crosslinker binding and unbinding dynamics, and their mechanical contribution.
In their resting state, the cells are flat ellipsoids and the MB is contained in a plane that is orthogonal to the minor cell axis. Assuming that the cell is discoid for simplicity ($R_1=R_2=R$) the major radius $R$ is also approximately the radius of the MB (Fig \ref{illus_intro}D), and thus the MTs bundled together in the MB have a curvature $\sim 1/R$. We first consider timescales larger than the dynamics of MT crosslinker binding and unbinding (about 10 seconds \cite{braun2011adaptive}), for which we can ignore the  mechanical contribution of crosslinkers \cite{bathe2008cytoskeletal}.
% We first assume we ignore MT crosslinker binding and unbinding dynamics and their mechanical contribution at long time scales.
Using the measured flexural rigidity $\kappa = 22\, pN \mu m^2$ of MTs \cite{gittes1993flexural}, and defining $\mathcal{L}$ as the sum  of all MTs length, the elastic energy of the MB is $E_B= \frac{\kappa}{2} {\mathcal{L}}/{R^2}$.
%A similar consideration holds for the cell cortex. 
At time scales larger than a few seconds, the cortex can reorganize and therefore we do not have to consider its rigidity \cite{salbreux2012actin}. Its effect can then be modeled by a surface energy associated with a surface tension $\sigma$  (Fig \ref{illus_intro}D).
The surface area is $S=2\pi R^2 [ 1+  O(\frac{r}{R}) ]$, in which $2r$ is the thickness of the cell. Assuming the cell to be flat enough, its surface area is therefore approximately $2 \pi R^2$ and the energy is $E_T \sim 2 \pi  \sigma R^2$. The equilibrium of the system corresponds to $\partial_R (E_B + E_T)=0$, leading to :
\be
\label{result_R}
R^4 = \frac{\kappa \mathcal{L} }{4 \pi \sigma} \, .
\ee
All other things constant, we thus expect $R \propto \mathcal{L}^{1/4}$. 
 To verify this relationship, we compiled data from 25 species available from the literature \cite{goniakowska1976evidence}, computing $\mathcal{L}$ by multiplying the number of microtubules in a cross-section by the length of the marginal band.
 The scaling is remarkably respected, over more than two orders of magnitudes (Fig. \ref{MB_radius}A).
Using equation \ref{result_R}, the fit provides an estimate of the tension of $\sigma \sim 0.1 pN/ \mu m$, which is low compared to the tension $\sigma \sim 100 pN/ \mu m$ of the actomyosin cortex of blebbing cell \cite{tinevez2009role}. However, RBC have a cortex made of spectrin rather than actomyosin, and thus have a much lower tension, that compensates a negative membrane tension \cite{fournier2004fluctuation}.  
In Human RBC, membrane tension was shown to be negative with a magnitude of $0.65\,pN/ \mu m$ \cite{turlier2016equilibrium}, close to the magnitude derived from our theory. 
In contrast to RBC, we predict  $\sigma = 40\,pN/ \mu m$ for Human blood platelets, given that $R \approx 2 \mu m$ and $\mathcal{L} \approx 100 \mu m$ \cite{kenney1985cystoskeleton}, which is close to the value reported for blood granulocytes ($35\,pN/ \mu m$) \cite{evans1989apparent}.

The precise scaling observed in the experimental data confirms our mechanically driven hypothesis where the MB pushes on the cell cortex, and in which at long time scales (on the order of a minute), only the bending rigidity of the MTs and the cortical tension need to be considered  (Fig \ref{illus_intro}D). 
To verify that this result is still valid for a ring of multiple dynamically crosslinked MTs, we developed a numerical model of cells with MBs in {\em Cytosim}, a cytoskeleton simulation engine \cite{nedelec2007collective}.
 {\em Cytosim} simulates stochastic binding/unbinding of connectors, and represents them by a Hookean spring between two MTs. 
For this work, we extended {\em Cytosim} to be able to model a contractile surface under tension that can be deformed by the MTs.
Cell shape is restricted to remain ellipsoidal, and is described by six parameters: three axes length $R_1$, $R_2$, $r$ and a rotation matrix, {\em i.e.} three angles.
The three lengths are constrained such that the volume of the ellipsoid remains constant. 
To implement confinement, any MT model-point located outside the cell is subject to inward-directed force $\mathbf{f}=k \mathbf{\delta}$, in which $\mathbf{\delta}$ is the shortest vector between the point and the surface and $k$ the confining stiffness.
Here for each force $\mathbf{f}$ applied on a MT, an opposite force $-\mathbf{f}$ is applied to the surface, in agreement to Newton's third law. The rates of change of the ellipsoid parameters are then given by the net force on each axis, divided by $\mu$, an effective viscosity parameter (see Suppl. 1.I.A). The value of $\mu$ affects the rate of cell shape change but not the stationary cell shape. This approach is much simpler than using a tessellated surface to represent the cell, and still general enough to capture the shape of blood platelets \cite{hartwig2002platelet,lee2004shape} and several RBCs \cite{schroter1990influence,cohen1991cytoskeletal}, see Fig. \ref{illus_intro}A. 

To model resting platelets, we simulated a ring made of  $10$--$20$ MTs of length $9$--$16 \mu m$ \cite{patel2008visualization} with $0$ or $10000$ crosslinkers, confined in a cell of volume $8.4 \mu m^3$ with a tension $\sigma \sim 0.45$--$45  pN/ \mu m$, for over six minutes, until equilibration. We find that the numerical results agree with the scaling law, over a very large range of parameter values as illustrated in Fig. \ref{MB_radius}B. 
Interestingly, we find that simulated cells are slightly larger than predicted analytically. This is because MTs of finite length  do not exactly follow the cell radius, and their ends are less curved, thus exerting more force on the cell. This means that the value of the tension we computed from the biological data ($\sigma \sim 0.1 pN/\mu m$) is slightly under-estimated. More importantly, the simulation shows that with or without crosslinkers, the cell has the same size at equilibrium (compare black and gray dots on Fig. \ref{MB_radius}B), confirming that, because they can freely reorganize, crosslinkers should not affect the long-term elasticity of the MB.
To understand the mechanics of blood cells with MBs at short time scale, however, it is necessary to consider the crosslinkers.

%These numerical results agree with the analytical prediction, over a wide range of parameters as illustrated in Fig. \ref{MB_radius}, B. 
%The fact that the results with and without crosslinkers nearly overlap also confirms that crosslinkers do not influence the stationary cell size. 
%The shape of the cell at equilibrium is thus solely determined by its cortical tension and the elasticity of the MTs.

%It shows in particular that at long time-scales, only the bending elasticity of the MTs and the cortical tension need to be considered. 
%We expect that at short time scale however, the molecules connecting MTs within the ring would play a more notable  role, and we investigated this effect with a numerical model of the platelet built with {\em Cytosim}, according to published methods \cite{nedelec2007collective}.

%\subsection*{Dynamic response of the MB to cell rounding}

\begin{figure}[t]
\includegraphics[width=0.45\textwidth]{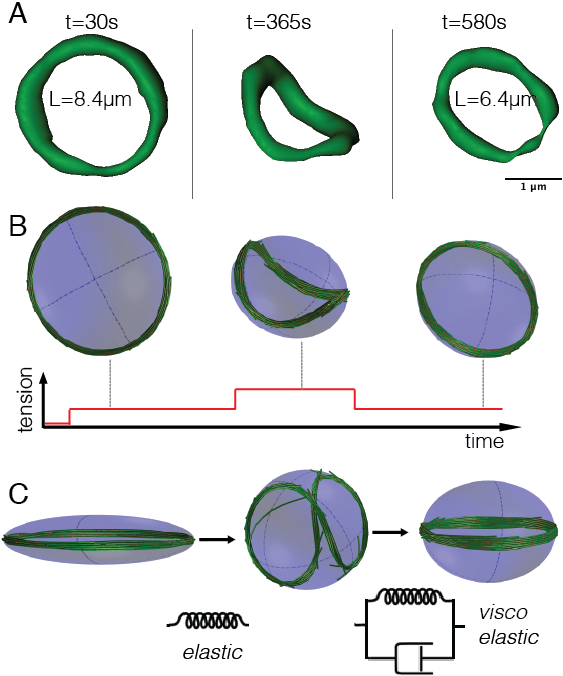}
\caption{
\label{illus_elasticity}
A) MB of a live platelet labeled with SIR-tubulin dye. Fluorescence images were segmented at the specified time after the addition of ADP, a platelet activator, to obtain the MB size $L$.
B) Simulation of a platelet at different times. A limited increase of the tension ($90pN / \mu m$) causes the MB to shorten while a large increase of the tension ($ 220 pN/\mu m$) causes the buckling of the MB. 
C) Simulations show that if cell rounding is fast enough, the MB buckles because crosslinkers cannot reorganize. This represent an elastic behavior, but at longer times, the MB rearranges, leading to a viscoelastic response.
}
\end{figure}

\subsection*{The marginal band behaves like a viscoelastic system}
During activation, mammalian platelets round up before spreading, and their MB coils during this process which occurs within a few seconds \cite{diagouraga2014motor}. Similar reports were made for thrombocytes \cite{lee2004shape}.
To observe these results experimentally, we extracted mice platelets, and activated them by exposing them to adenosine diphosphate ADP, causing an often reversible response. By monitoring the MB with SiR-tubulin, a bright docytaxe-based MT dye, we could capture the MB coiling live, Fig. \ref{illus_elasticity}A.
As it coils, the MB adopts the shape of the baseball seam curve, which is the shape that an incompressible elastic ring would adopt when constrained into a sphere smaller than its natural radius \cite{guven2012confinement}. 
Thus, at short time scale, the MB seems to behave as an incompressible ring, and we reasoned that this must be because crosslinkers prevent MTs from sliding relative to each other. 
To analyse this process further, we returned to {\em Cytosim}. After an initialisation time, in which the MB assembles as a ring of MTs connected by crosslinkers, cortical tension is increased stepwise. The cell as a consequence becomes spherical, and, because its volume is conserved, the largest radius is reduced compared to that of the discoid resting state. As a result, the MB adopts a baseball seam shape (Fig. \ref{illus_elasticity}B). 
Over a longer period, however, the MB regained a flat shape, as MTs rearranged into a new, smaller, ring (Fig. \ref{illus_elasticity}A).
In conclusion, the simulated MB is viscoelastic (Fig. \ref{illus_elasticity}B). At short time scales, MTs do not have time to slide, and the MB behaves as an incompressible elastic ring. At long time scales, the MB behaves as if crosslinkers were not present, with an overall elastic energy that is the sum of individual MT energies. 
Thus overall, the ring seems to transition from a purely elastic at short time scales, to a viscoelastic Kelvin-Voigt law at long time scales (Fig. \ref{illus_elasticity}C). The transition between the two regimes is determined by the timescale at which crosslinkers permit MTs to slide.
% We can estimate the effective viscosity of the microtubule bundle as $\mu_{MB} \sim k_x N \tau_{off}$ in which $k_x$ is the elasticity of the crosslinkers, $N$ is the number of bound crosslinkers, and $\tau_{off}$ is the unbidning rate of crosslinkers. Coiling should thus occur when $\mu_{MB} \gg \mu$.

\subsection*{The cell is unexpectedly robust}

\begin{figure}[t]
\includegraphics[width=0.45\textwidth]{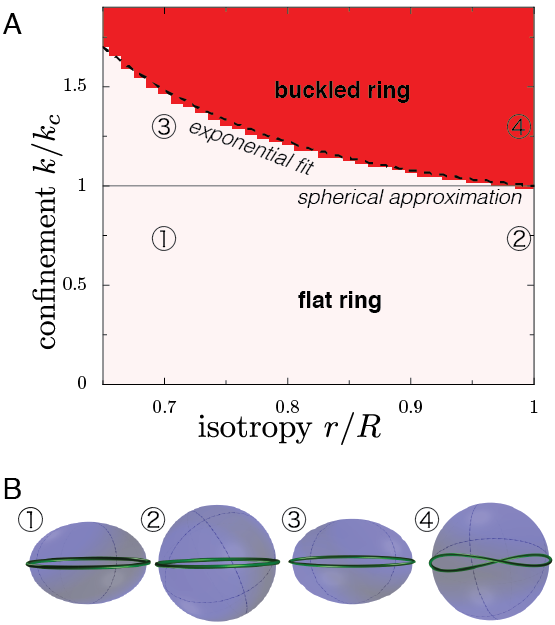}
\caption{
\label{pd_conf_geom}
A) Degree of coiling as a function of normalized confinement stiffness $k/k_c$ and isotropy $r/R$ of the fixed oblate ellipsoid in which the ring is confined. The light shade indicates regions of uncoiled states and the red area indicates coiled states, as determined by simulations. The dashed line represents the empirical function $k^*=k_c(\frac{r}{R})^2 e^{\alpha (1-\frac{R}{r})}$, where $\alpha=2.587$ is a phenomenological parameter that depends on $\epsilon$, itself defined from the MB length as $L=2 \pi R (1+\epsilon)$. B) Illustrations of MB shapes in different regimes, as indicated by the circled numbers.
}
\end{figure}

The MB in blood cells is necessary to establish a flat morphology, but also to maintain this morphology in face of transient mechanical challenges, for example as the cell passes through a narrow capillary \cite{joseph1984cytoskeletal}.
In this section, we calculate the response of a cell to a fast mechanical stimuli during which crosslinkers do not reorganize.
Therefore, we can assume that the ring is uniform and of constant length, to investigate how cortical tension affects the resistance of the cell to coiling.
Firstly, we examine the mechanics of a closed ring of length $L$ and rigidity $\kappa_r$ within a sphere, and then extend these results to a non-deformable ellipsoid. The shape of a ring in a sphere was previously calculated numerically \cite{guven2012confinement}, and we extended these result by deriving analytically the force $f_B$ required to buckle a confined ring (see Suppl. 1.II.B). 
If $E_{B}$ is the energy of a buckled MB, the force is :
\be
f_B = - \lim_{L\rightarrow 2 \pi R} \partial_R E_{B} = 8 \pi \frac{\kappa_r}{R^2}
\label{buckling_force}
\ee
We verified this relation in simulations, with $L=2 \pi R (1+\epsilon)$, where $1 \gg \epsilon > 0$, which made the ring slightly oversized compared to its confinement. Given the confining stiffness $k$, the force applied to each model-point of the ring is $k R \epsilon$. If $n$ is the number of model-points in the rings ({\it i.e.} $n=L/s$ where $s$ is the segmentation), the total centripetal force is $n k R \epsilon$. Hence, we expect that the ring will buckle if $k$ exceeds $k_c= \frac{1}{n R \epsilon} f_B$.
Upon systematically varying $k$ in the simulation (see Methods), we indeed found that the ring coils for $k>k_c$, Fig. \ref{pd_conf_geom}A.
We next simulated oblate ellipsoidal cells, with $R_1=R_2=R$ and $r<R$, and we varied the flatness of the cell by changing $r/R$. We found that the measured critical confinement $k^*$ is indeed $k_c$ for $r=R$, but increases exponentially with $1-R/r$, Fig. \ref{pd_conf_geom}. The buckling force of a MB is thus much higher when the MB is confined.
This is important mechanically, as it implies that the flat state of the MB should be metastable, and this  could make a blood platelet 50 times more resilient to buckling (assuming an isotropy ratio $r/R=0.25$).

%It is interesting to remark that the buckling force of a ring (Eq. \ref{buckling_force}) is a hundred time greater than the Euler force threshold for buckling of a linear bundle ($\pi^2 \kappa_r / L^2$).
%Thus if the cell is flat, the cortical tension reinforces the MB laterally, and the overall cell rigidity is enhanced, making it strikingly resistant to buckling.
%Thus flatter cells are much stronger at resisting external forces.

\subsection*{Coiling stems from  cortical tension overcoming MB rigidity}

\begin{figure}[t]
\includegraphics[width=0.4\textwidth]{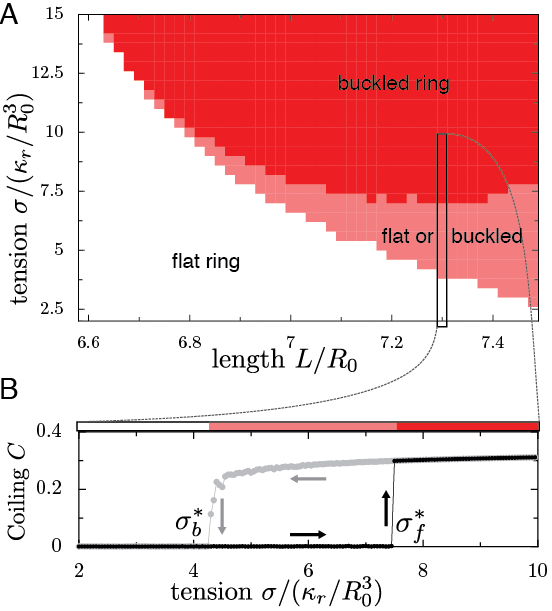} 
\caption{
\label{phase_diags}
A) Configuration of the MB as a function of renormalized tension $ \sigma R_0^3 /\kappa_r$ and renormalized MB length $ L/R_0$, in which the volume of the cell is $\frac{4}{3} \pi R_0^3$. 
The state of the MB is indicated by colours: gray: flat MB ; red : always buckled MB ; pink : bistable region, in which the MB can be either buckled or flat.
B) A cut through the phase diagram, for a MB of length $L=7.3 R_0$. The degree of coiling (see methods for definition) as a function of tension, in a cell initially flat (black dots) or buckled (gray dots), shows the metastability of the flat state. Arrows illustrate the hysteresis.
}
\end{figure}

We can now consider the case of a ring inside a deformable ellipsoid of constant volume $V_0 = 4/3 \pi R_0^3$, governed by a surface tension $\sigma$. 
The length of the ring $L$ is set with $L > 2\pi R_0$, such that we expect the ring to remain flat, at low tension, and to be coiled, at high tension, because it does not fit in the sphere of radius $R_0$.
%Since we observed exactly this behaviour, we could extract the critical tension as a function of other parameters of the simulations (Fig. \ref{phase_diags}A).
%As discussed in the previous section, the critical tension is higher than can be predicted theoretically for a ring in a sphere [CITE], because the cell is ellipsoidal rather than spherical.
%In simulations, we observe as expected the existence of a critical tension leading to coiling, Fig. \ref{phase_diags}A. This shows that increasing $\sigma R_0^3 / \kappa_r$, i.e. increasing the ratio of cortical tension over MB rigidity, leads to cell rounding. Thus, either increasing the cortical tension or destabilizing the MB will lead to coiling.  Moreover, the MB in the contractile ellipsoid exhibits bistability and hysteresis : a cell initially flat will remain flat above the critical tension, while a cell initially round will remain round even below the critical tension, Fig. \ref{phase_diags}A,B. As we could expect, the flat configuration is metastable because a MB in a flat cell has a higher buckling threshold than a MB in a spherical cell, as discussed in the previous section.
In simulations, starting from a flat ring, we observe as predicted the existence of a critical tension $\sigma_f^*$ leading to buckling, Fig. \ref{phase_diags}A. This shows that increasing $\sigma R_0^3 / \kappa_r$, {\it i.e.} increasing the ratio of cortical tension over ring rigidity, leads to cell rounding. Thus, either increasing the cortical tension or weakening the ring will lead to coiling. Starting from a buckled ring, decreasing the tension below a critical tension $\sigma_b^*$ also leads to the cell flattening, as predicted.
However, our simulations show that $\sigma_b^* < \sigma_f^*$ : a cell initially flat will remain flat for $\sigma_b^* < \sigma <\sigma_f^*$, while a cell initially round will remain round for $\sigma_b^* < \sigma <\sigma_f^*$, Fig. \ref{phase_diags}A,B. Hysteresis is the hallmark of bistability, and we had predicted this bistability in the previous section by showing that the flat configuration is metastable. This metastability, {\em i.e.} the fact that a MB in a flat cell has a higher buckling threshold than in a spherical cell, allows the cell to withstand very large mechanical constraints such as shear stresses.

%, since a MB in a flat cell has a higher buckling threshold than a ring in a spherical cell.

\section*{Discussion}

We have examined how the forces determining the morphology of blood cells balance each other.
In particular, we predicted a scaling law $4 \pi R^4 = \kappa \mathcal{L} / \sigma$, if the elasticity of MTs is compensated by cortical tension, in which $\mathcal{L}$ is the sum of the lengths of the MTs inside the cell, $\kappa$ the bending rigidity of MTs and $\sigma$ the cortical tension. 
Remarkably, this scaling law is well respected by values of $R$ and $\mathcal{L}$ measured for 25 species. 
%Unless $\kappa$ was variable, which would be extraordinary surprising considering the conservation of basic cytoskeletal components, this indicates that the tension is relatively uniform across these species.
We caution that these observations were made for non-discoidal RBC (where the two major axes differ), indicating that other factors not considered here must be at work \cite{joseph1984cytoskeletal}.
In human RBC, perturbation of the spectrin meshwork can lead to elliptical RBC \cite{lux1978diminished}, showing that the cortex can impose anisotropic tensions, while another study suggests that MB-associated actin can sequester the MB into an elliptical shape \cite{cohen1998elliptical}. Cortical anisotropy would be an exciting topic for future studies, but this may not be needed to understand wild-type mammalian platelets.

Using analytical theory and numerical simulations, we analyzed the mechanical response of cells with MB, and showed a complex viscoelastic behavior characterized by a timescale $\tau_c$ that is determined by crosslinker reorganization.
At long time scales ($t\gg\tau_c$), the MB behaves elastically, and its elasticity is the sum of all MTs rigidity.
At short time scales ($t< \tau_c$), the MB behaves as an incompressible elastic ring of fixed length because crosslinkers do not yield. At this time scale, the stiffness of the ring exceeds the sum of the individual MT stiffness as long as the crosslinkers connect neighboring MT tightly \cite{ward2014mechanical}.
Buckling leads to the baseball seam curve, which is a configuration of minimum elastic energy.
This explains the coiled shape of the MB observed in mouse platelets, as well human platelets \cite{diagouraga2014motor} and well as dogfish thrombocytes \cite{lee2004shape}. 
%The viscous timescale is determined by the crosslinkers, and the ability of MT to slide relative to each other in the bundle.
Thus an increase of cortical tension over bundle rigidity can cause coiling, if the cell deforms faster than the MB can reorganize. 
A fast increase of tension is a likely mechanism supported by several experimental evidence \cite{hartwig1992mechanisms,carroll1982phosphorylation,li2002arp2}. In dogfish thrombocytes and platelets, blebs are concomitant with MB coiling, suggesting a strong increase of cortical tension \cite{lee2004shape}. 
We note however that a recent study suggests that MB destabilization could be due to ring extension \cite{diagouraga2014motor}.

Finally, calculating the buckling force of a cell containing an elastic MB and a contractile cortex led to a surprising result.
We found that the buckling force increased exponentially with the cell flatness, because the cortex reinforces the ring laterally. This makes the marginal band a particularly efficient system to maintain the structural integrity of blood cells. For transient mechanical constraints, the MB behaves elastically and the flat shape is metastable, allowing the cell to overcome large forces without deformation. However, as we observed, the viscoelasticity of the MB allows the cell  to adapt its shape when constraints are applied over long timescales, exceeding the time necessary for MB remodeling by crosslinker binding and unbinding.
This study suggest that it will be particularly interesting to compare the time-scale at which blood cells experience mechanical stimulations {\it in vivo}, with the time scale determined by the dynamics of the MT crosslinkers.

%This indicates that having a MB is mechanically particularly advantageous to these type of cells that exhibit a very flat morphology and are subjected to large shearing stresses.

%\bibliography{BibTeX_ref_pio}

\section*{Methods}

MTs of persistence length $l_p$ are described as bendable filaments of rigidity $\kappa = k_B T l_p$, in which $k_B T$ is the thermal energy. We can write the energy of such a filament of length $L$ as the integral of its curvature squared :
\be 
\label{lagrangian}
E= \frac{\kappa}{2}\int_0^L  \left(\frac{d^2\mathbf{r}}{ds^2}\right)^2\,  ds
\ee
Where $\mathbf{r}(s)$ is the position as a function of the arclength $s$ along the filament. The dynamics of such a system was simulated in Cytosim, an Open Source simulation software \cite{nedelec2007collective}. 
In Cytosim, a filament is represented by model points distributed regularly defining segments of length $s$. 
Fibers are confined inside a convex region of space $\Omega$ by adding a force to every model points that is outside $\Omega$.
The force is $\mathbf{f}=k (\mathbf{p}-\mathbf{r})$, where $\mathbf{p}$ is the projection of the model point $\mathbf{r}$ on the edge of $\Omega$. 
For this work, we implemented a deformable elliptical surface confining the MTs, parametrized by six parameters. The evolution of these parameters is implemented using an effective viscosity (see Suppl. 1.I.C).
To verify the accuracy of our approach, we first simulated a straight elastic filament, which would buckle when submitted to a force exceeding $\pi^2 \kappa / L^2$, as shown by Euler. Cytosim recovered this result numerically.
For a closed circular ring, we also find that the critical tension necessary for buckling corresponds very precisely to the theoretical prediction \cite{lee2007compressing}.
This is also true for an elastic ring confined inside a prolate ellipsoid of tension $\sigma$ (see Suppl. 1.I.D). 

To simulate cell radius as a function of $\mathcal{L} \kappa / \sigma$, we used a volume of $8\pi/3 \mu m^3$ (close to the volume of a platelet), with a tension $\sigma \sim 0.45$--$45  pN/ \mu m$, consistent with physiological values. The MTs have a rigidity $22\, pN\,\mu m^2$ as measured experimentally \cite{gittes1993flexural}. We simulate $10-20$ MTs of length $9-16 \mu m$, and with a segmentation of $125nm$, we used more than 70 points per MTs. The crosslinkers have a resting length of $40 nm$, a stiffness of $91  pN/\mu m$, a binding rate of $10 s^{-1}$, a binding range of $50 nm$, and an unbinding rate of $6 s^{-1}$. An example of simulation configuration file is provided in Suppl. 2.
When considering an incompressible elastic ring, we used a cell of volume $4/3 \pi R_0^3$, where $R_0$ is the radius of the resting (spherical) cell. For simplicity, we can renormalize all lengths by $R_0$ and thus all energies by $\kappa_R/R_0$.  We simulate a cell with a tension $\sigma = 5-18 \kappa_R/R_0^3$, and a ring of length $1-1.2 \times 2 \pi R_0$.
To test the effect of confinement, we place an elastic ring of rigidity $\kappa$ in an ellipsoid space of radii $R_0,\, R_0,\, r\,R_0$, in which $r<1$. The elastic ring has a length $(1+\epsilon) 2 \pi R_0$, in which $\epsilon =0.05$.
To describe how coiled is a MB, we first perform a principal component analysis using all the MTs model points. The vector $\mathbf{u}_z$ is then set in the direction of the smallest eigenvalue while $\mathbf{u}_x,\mathbf{u}_y$ are set orthogonally. We can then define the degree of coiling $C$ as the deviation in $Z$ divided by the deviations in $XY$:
\be
C = \sqrt{ \frac{\sum z^2 }{\sum x^2 + y^2}}
\ee
Thus, $C$ is independent of the size of the cell and only measures the deformation of the MB. 
%To show that our simulation is accurate, we simulate a  MT bundle  inside an ellipsoid cell of tension $\sigma$ and viscosity $\mu$. In the {\em Methods} section, we show that below a critical tension $\sigma^*$, the filament will extend the space into a prolate ellipsoid, while for $\sigma>\sigma^*$, the tension will buckle the filament, a phenomenon called {\em Euler buckling}. We can simulate this system for a range of filament length $L$ and tension $\sigma$, and we find that our simulations match very accurately the analytical solution for the value of $\sigma^*$.
To measure the critical value of a parameter $\mu$ (e.g. tension or confinement) leading to coiling, we computed the derivative of the degree of coiling $C$ with respect to this parameter. Because buckling is analogous to a first-order transition, the critical value $\mu^*$ can be defined by :
\be
\partial_\mu C \big|_{\mu^*} = \max{\| \partial_\mu C \|} 
\ee
%{\color{red} We use this to measure the critical confinement $k^*$ and the critical tensions $\sigma_b^*$ and $\sigma_f^*$.}
Platelets were extracted using a previously published protocol \cite{cazenave2004preparation}, and labeled by SiR-tubulin \cite{lukinavivcius2014fluorogenic} purchased from Spirochrome.

\section{SI I : Simulation of microtubules/cortex interaction}
To understand cell shape maintenance, one needs to model the interaction between the cellular cortex and the microtubule marginal band. The structure of the marginal band is well known, compared to the organization of the cortex which is not caracterized. We thus decided to represent the microtubules individually, and the cortex effectively as a contractile surface. The interactions between a discretized ({\it e.g.} triangulated) surface and discrete filaments can be a demanding problem computationally, since such a surface would have a very large number of degrees of freedom. In contrast, we describe here how the problem remains relatively simple for a continuous shape that is described by a limited number of parameters. 

\subsection{General formulation}
\subsubsection{Forces and parametrization}
Let $S(p_k)$ be the surface defined by the set of parameters $\{p_k\}_{k<n}$. Let $\{\mathbf{f}_i\}_{i<m}$ be the set of forces applied on $S$ at the points  $\{\mathbf{r}_i\}_{i<m}$. They are defined by $-\mathbf{f}_i=\partial E / \partial \mathbf{r}_i$, where $E$ is the energy of the system (excluding the surface). One can define ``effective forces''  $\{\phi^k\}_{k<n}$ associated with each degree of freedom of the surface :
\begin{equation}
\phi^k = -\frac{\partial E}{\partial p_k} = \sum_{i<m} \mathbf{f}_i . \frac{\partial \mathbf{r}_i}{\partial p_k}  \label{pseudofor}
\end{equation}
We can define $\delta E$ the infinitesimal change in energy after an infinitesimal set of displacements $\delta \mathbf{r}_i$, and then compute it as a function of the infinitesimal set of parameter changes $\delta p_k$.
\begin{eqnarray}
\delta E= - \sum_i \delta \mathbf{r}_i . \mathbf{f}_i \\
\delta \mathbf{r_i} = \sum_k \delta p_k \frac{\partial \mathbf{r}_i}{ \partial p_k} \label{displacements} \\
\delta E =- \sum_k \delta p_k \phi^k \label{energies}
\end{eqnarray}
To write equation \ref{displacements}, we had to assume that any displacement of the surface (allowed by the constraints) can be described in terms of $p_k$, i.e.  that $S(\{p_k\})$ is surjective. It is here interesting to notes that $\phi^k$ has the dimension of a force if $p_k$ is a length, while it has the dimention of a torque (i.e. an energy) is $p_k$ is an angle.

\subsubsection{Constraints}
In many cases, constraints can be introduced using Lagrange multipliers, by inserting them into the energy $E$. For instance, to maintain the volume, we can define an energy $E'=E+PV$ where $V$ is the volume and $P$ is the pression ; here $P$ is also a Lagrange multiplier and we have to calculate its value appropriately to obtain $V=V_0$. The  pseudo-forces $\phi^k_P$ associated to pressure are :
\begin{eqnarray}
\phi^k_P=- P \frac{\partial V}{\partial p_k}
\end{eqnarray}

%The marginal band can be modeled numerically as a set of microtubules with crosslinkers and bundlers since it has a fairly low number of filaments, and its organisation can be observed in electron microscopy. On the other hand, the cortex architecture is much less characterized and is, in the case of red blood cells, a complex assembly of spectrin with a large number of heavily connected short actin filaments, making it very challenging to simulate in all its details. However, the cortex is mostly continuous and has been modeled, with great success, as a continuous sheet. This sheet undergoes a tension due to actin contractility, while resisting deformations as a visco-elastic body.

%Simulating the interactions between discrete microtubules and a continuous cortex is arduous problem as well,  as most numerical tools deal with discrete/discrete or continuous/continuous interactions. However, we show in the methods section that this problem becomes much simpler if the continuous shape can be described by a limited number of parameters. Blood  platelets have  a strikingly ellipsoidal shape, in agreement with cell shape models. We can thus assume the cell cortex to take an ellipsoidal shape, and we can thus simulate its interactions with the marginal band.

\subsection{Contractile Ellipsoid}
In this section, we describe a more complex, 3D surface. We model an ellipsoid centered around $0$ that has a fixed volume $V_0$ and a a surface tension $\sigma$, which means an energy $E_n = \sigma S$, where $S$ is the surface area of the ellipsoid. The ellipsoid is described by its eigenvectors $\mathbf{u}_{1,2,3}$ and their eigenvalues (i.e. the radii of the ellipse) $a_{1,2,3}$. We will also use the orientation matrix $M=[\mathbf{u}_{1},\mathbf{u}_{2},\mathbf{u}_{3}]$.
By construction, $M$ is a rotation matrix of determinant 1.

\subsubsection{Surface Tension}
We can compute the pseudo-forces associated to surface tension as:
\begin{eqnarray}
%\phi_P^1 = \frac{4}{3} \pi P r_2 r_3 \, ; \,
%\phi_P^2 = \frac{4}{3} \pi P r_3 r_1 \, ;  \,
\phi_\sigma^k = - \sigma \frac{\partial S}{\partial a_k}
\end{eqnarray}
The surface area of an ellipsoid is a complex special function that is not a combination of the usual functions. For convenience, we used an analytical approximation of the area : 
\begin{equation}
S(a_1,a_2,a_3)\; \simeq \; 4 \pi \left(\frac{(a_1 a_2)^p+(a_3 a_2)^p+(a_1 a_3)^p}{3}\right)^\frac{1}{p}
\label{surf_ell}
\end{equation}
For which $p=1.6075$ is an empirical parameter. This formula yields an error usually below a percent. 

\subsubsection{Point Forces}
To add the contribution of the external forces exerted by the  microtubules on the surface, we need to determine $\partial \mathbf{r} / \partial p_k$. The position of a point on the surface of the ellipsoid is defined by two angles $\theta$, $\phi$ as :
\begin{eqnarray}
\mathbf{r}=\mathbf{u}_1 r_1 \cos{\theta} \sin{\phi} + \mathbf{u}_2 r_2 \sin{\theta} \sin{\phi} + \mathbf{u}_3 r_3 \cos{\phi} 
\end{eqnarray}
Therefore, we have :
\begin{eqnarray}
\frac{\partial \mathbf{r}}{\partial r_k} = \mathbf{r} . \mathbf{u}_k
\end{eqnarray}
It is clear that only the component of the force normal to the surface is providing work upon changing $r_k$, therefore, we can discard the tangential component when computing the radial forces.
 The contribution of a force $\mathbf{f}$ at a point $r$ with a local normal $\mathbf{n}$ to the pseudo force $\phi_f^k$ is therefore :
\begin{eqnarray}
\phi_f^k = f_n \frac{\mathbf{n} . \mathbf{u}_k}{\| \mathbf{u}_k \|} \\
\text{With : } f_n=\mathbf{f}.\mathbf{n} 
\end{eqnarray}

We can now compute the torque generated by $\mathbf{f}$. In 2D, it would be convenient to describe the ellipse orientation by an angle $\theta$, and the result is that the  "angular force"  $\phi^\theta$ is the torque $\mathbf{r} \times \mathbf{f}$. We will assume that this is general and stays true in 3D ; thus we can write $\phi^{ang}$ directly as a vector :
\begin{eqnarray}
\boldsymbol{\phi}^{ang} = \mathbf{r} \times \mathbf{f}
\end{eqnarray}

\paragraph{Volume conservation}
To implement volume conservation we only need to find a pressure $P$ such that $(V-V_0)/V_0<\epsilon$ where $\epsilon$ is the tolerance. Many techniques allow the convergence to a suitable value of $P$ - and the choice of method has no physical implication. Here we used a gradient descent method known as the shooting method. For this, we start with an initial value of $P=0$ and we compute $V(P)$. If $V(P)-V_0 / V_0 > \epsilon$, we compute $V(P+\delta P)$ to get the gradient of the volume with respect to pressure. We then follow this gradient until we reach the desired aim for $V$. This method works very well if $V(P)$ is monotonous, which is always the case here.

The volume of the ellipse is $V=\frac{4}{3} \pi a_1 a_2 a_3$ and therefore, using the Lagrange multiplier $P$ to conserve the volume we can write :
\begin{eqnarray}
%\phi_P^1 = \frac{4}{3} \pi P r_2 r_3 \, ; \,
%\phi_P^2 = \frac{4}{3} \pi P r_3 r_1 \, ;  \,
\phi_P^k = \frac{4}{3} \pi P \frac{a_1 a_2 a_3}{a_k}
\end{eqnarray}

\subsection{Time Evolution}

We can now define the time evolution of the ellipse. We assume a unique viscosity $\mu$ associated to the change of size of the ellipse, and a rotational viscosity $\eta_{ang}$.
\begin{eqnarray}
\dot{a_k} = \frac{1}{\mu}\left(\phi_P^k + \phi_\sigma^k+\sum \phi_{f}^k\right) \\
\dot{M} = R \left( \mathbf{u} \right) \quad \text{with} \quad \mathbf{u}=\frac{1}{\eta_{ang}} \sum \boldsymbol{\phi}^{ang}_{ f}
\end{eqnarray}
In which $R(\mathbf{u})$ is the rotation matrix generated from the moment vector $\mathbf{u}$.

\subsection{Simulations Validation}

To validate out numerical method and its implementation, we first simulated a microtubule bundle confined inside an ellipsoid cell of tension $\sigma$ and volume $\frac{4}{3} \pi R_0^3$. A classical result of analytical mechanics is that a filament should buckle under a force tangential force $f$ is this force is larger than a critical force :
\begin{equation}
f_1^* = \frac{\kappa \pi^2}{L^2}
\end{equation}
Assuming the microtubules to be sliding freely, the critical buckling force of a  microtubules is thus $f_n^*=n f_1$. We confined the microtubule in a contractile ellipsoid, which thus takes the shape of a prolate ellipsoid.  Let us call $a_1 = R$ the longer axis of this ellipsoid, and the shorter axis are $a_2,a_3 = \sqrt{R_0^3 / R}$. The force exerted on the microtubule is $f_\sigma = 2 \sigma \partial_R S(R, \sqrt{R_0^3 / R}, \sqrt{R_0^3 / R})$, with $S$ defined in equation \ref{surf_ell}. Starting with a microtubule of length $L$, buckling will occur for a critical tension :
\begin{equation}
\label{buckling_bundle}
\sigma^* = \frac{\kappa \pi^2}{2 R^2  } \left( \partial_R S(R, \sqrt{R_0^3 / R}, \sqrt{R_0^3 / R})\right)^{-1}
\end{equation}

%In the {\em Methods} section, we show that below a critical tension $\sigma^*$, the filament will extend the space into a prolate ellipsoid, while for $\sigma>\sigma^*$, the tension will buckle the filament, a phenomenon called {\em Euler buckling}. We can simulate this system for a range of filament length $L$ and tension $\sigma$, and we find that our simulations match very accurately the analytical solution for the value of $\sigma^*$.

\begin{figure}
                \includegraphics[width=0.4\textwidth]{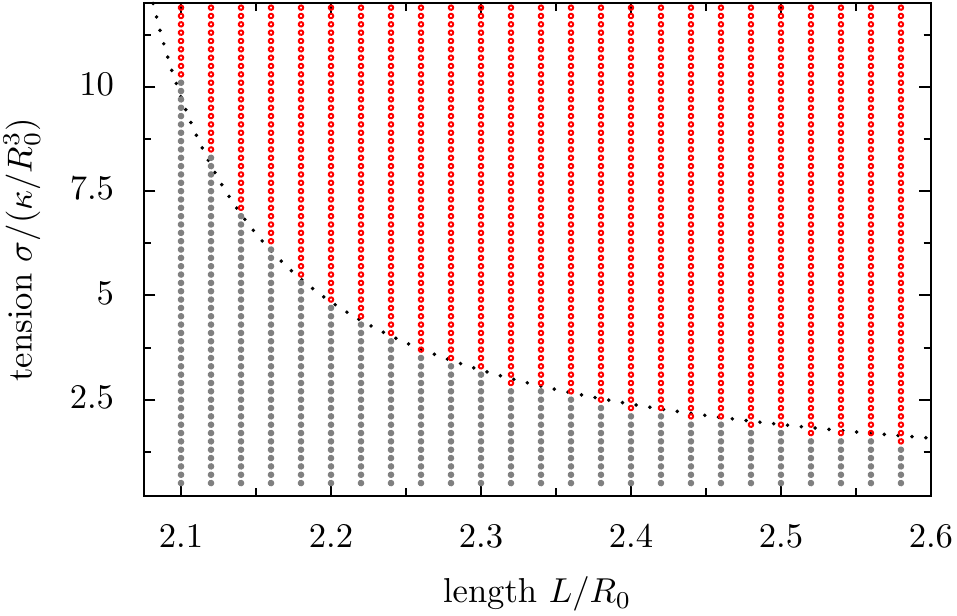}
                \caption{Phase diagram of the degree of buckling as a function of the length and the tension. Red means that the filament is buckled, gray that it is flat. The dashed line represents the critical tension calculated in Eq. \ref{buckling_bundle}.}
                \label{Ellipsoid}
\end{figure}
%The cortex has $\sigma$, and a viscosity $\mu$ Additionally, we assume the  cell volume $V$ to be fixed, in agreement with the existence of mechanism for cells to maintain their physiological volume. The default shape of a such cell with tension and conserved volume is a sphere but point forces, in particular coming from cytoskeletal filaments, can deform the cell surface.  Since the ellipsoid is the first mode of deformation of the sphere, our approach is also analogous to an expansion around the spherical cell shape. 

\section{SI II : Mechanics of a confined ring}

\subsection{Formulation}
Let us consider a rod of length $L$ lying on a sphere of radius $R$. We can describe this rod by its position $\mathbf{r}$, parametrized by its arclength $s$, such that the energy reads :
\begin{equation}
\label{energy_base}
E(R,L)=\frac{\kappa}{2}\int_0^L \ddot{\mathbf{r}}^2 ds
\end{equation}

Because the rod lies on the unit sphere, and because $s$ is the arclength, we have the constraints :
\begin{equation}
\| \mathbf{r} \|^2  = R^2
 \quad\quad \text{and} \quad\quad
\| \dot{\mathbf{r}} \|^2 =1 
\end{equation}
We can introduce this as constraints in the energy using two Lagrange multipliers $\alpha$ and $\beta$, to define :
\begin{equation} 
\label{lagrangian2}
E=\frac{\kappa}{2}\int_0^L \left[ \ddot{\mathbf{r}}^2 + \alpha (R^2- \| \mathbf{r} \|^2 ) +\beta(\| \dot{\mathbf{r}} \|^2  -1) \right] ds
\end{equation}
Minimizing this energy yields the Euler-Lagrange equation : 
\begin{equation}
\label{eulerlagrange}
\mathbf{r}^{(4)} = \alpha \mathbf{r} + \dot{\beta} \dot{\mathbf{r}} + \beta \ddot{\mathbf{r}}
\end{equation}

Since the curve is lying on a sphere, we can use the identity :
\begin{equation}
\label{seamdiffeq}
\ddot{\mathbf{r} }(s) = k_i(s) \left[ \mathbf{r}(s) \times \dot{\mathbf{r} }(s) \right] - \frac{1}{R^2} \mathbf{r} (s)
\end{equation}
In which $k(i)$ is the intrinsic (geodesic) curvature.  Eventually, we find :
\begin{equation}
\label{diffeqres}
\ddot{k_i} = \frac{k_i}{R^2} \left( \gamma - \frac{R^4}{2} k_i^2 \right) 
\end{equation}
In which $\gamma$ is a constant \cite{guven2012confinement}. To find the shape of a closed ring, one needs to find the value of $\alpha$ and $k_i(0)$ such that the curve is of length $L$ is a closed ring, i.e. :
\begin{equation}
\mathbf{r}(L)=\mathbf{r}(0) \quad \text{and} \quad \dot{\mathbf{r}}(L)=\dot{\mathbf{r}}(0)
\end{equation}
Numerically, we determined $\gamma$ and $k_i(0)$ using a shooting method. 

\subsection{Weakly deformed ring}
For a weakly deformed ring, equation \ref{diffeqres} can be simplified to 
\begin{equation}
\label{diffeqsimp}
\ddot{k_i} = \frac{\gamma}{R^2} k_i
\end{equation}

Periodicity imposes $\sqrt{-\gamma} \rightarrow m$ when $L\rightarrow 2\pi R$, in which $m \in \mathbb{N}$. Since the lowest energy curve has a period $L/2$, we can conclude that $m=2$, i.e. $\gamma \rightarrow - 4$ for $L \rightarrow 2\pi R$. 
Although analytically solving the full shape equation $\mathbf{R}(s)$ is arduous \cite{ostermeir2010buckling} even in this weakly deformed approximation, we can construct a shape equation that satisfies Eq. \ref{diffeqsimp} for small deformation as follows:
\begin{equation}
\mathbf{R} = R \colvec{3}{(1-b) \sin{t} + b \sin{3t} }{(1-b) \cos{t} -b \sin{3t}}{2\sqrt{(1-b)b} \cos{2 t}} \label{ersatz}
\end{equation}
In which $0\le t \le 2 \pi$ is a angle coordinate. For small deformations $b\rightarrow 0$, one finds :
\begin{eqnarray}
k_i (s)=6 \sqrt{b} \cos{2 s/R} + O\left( b^\frac{3}{2} \right) \\
\ddot{k_i}(s) = \gamma \times 6 \sqrt{b} \cos{2 s/R} +  O\left( b^\frac{3}{2} \right) ,
\end{eqnarray}
with $\gamma=-4$ as expected.  From equation \ref{ersatz}, we can compute the bending energy of the marginal band :
\begin{equation}
E(R,b)= \int_0^{2 \pi} \left( \frac{1}{R^2}+k_i^2  \right) \| \partial_t \mathbf{R} \|^2 dt
\end{equation}
For small deformations $b\rightarrow 0$, we have :
\begin{equation}
E(R,b) =\frac{\kappa}{2R }\left( 2 \pi + 36 \pi  b + O\left( b^2 \right) \right)
\end{equation}

\begin{figure}[t]
                \includegraphics[width=0.4\textwidth]{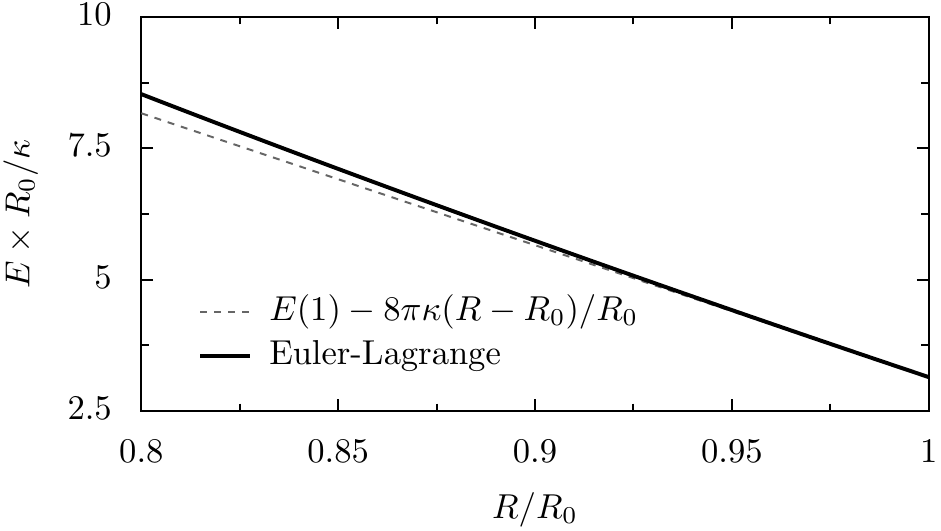}
                \caption{Bending energy of an incompressible elastic ring of length $2\pi R_0$ (the marginal band) in a sphere of radius $R<R_0$. The solid line represents the numerical solution to the Euler-Lagrange equations (Eq. \ref{diffeqres}), while the dashed line represents the small deformation approximation, Eq. \ref{force_app}. }
                \label{EofR}
\end{figure}

We can also compute the length of the marginal band, and the energy :
\begin{eqnarray}
L(R,b) =2 \pi R \left( 1 + 6  b + O\left( b^2 \right) \right),\\
E(R,L) \rightarrow \frac{\kappa}{2R} \left( 2 \pi + 3 \frac{L-2\pi R}{R} \right)
\end{eqnarray}

We then find the force exerted by a nearly flat ring on the sphere $L=2 \pi R$:
\begin{equation}
f_B= \lim_{L\rightarrow 2 \pi R} \partial_R E(R,L) = \frac{8 \pi \kappa}{R^2}
\label{force_app}
\end{equation}
This result is in agreement with solving the full shape equation (Eq. \ref{diffeqres}), as illustrated in figure \ref{EofR}. $f_b$ is the force exerted by a nearly flat ring on a sphere ; by construction it is also the critical force at which a ring will buckle. Numerically, we can study ring buckling in two cases : when the ring is undergoing an elastic confinement, and when the ring is confined by a contractile surface.

\subsection{Ring under elastic confinement}
  Let us consider a ring of length $L$ confined in a a sphere of radius $R$ such that $L=2 \pi (R + \epsilon)$ by an elastic confinement $k$ (see \cite{nedelec2007collective} for the implementation of confinement). The confinement force is here $f_c = k n \epsilon$, in which $n$ is the number of points describing the discrete ring. The ring will buckle if $f_c > f_B$ ; using Eq. \ref{force_app}, we find that the ring will buckle above a critical confinement : 
\begin{eqnarray}
k_c = \frac{8 \pi \kappa}{n  \epsilon R^2},
\end{eqnarray}
in which $n=L/ds$, where $ds$ is the segmentation.

\subsection{Ring in a contractile ellipsoid}
Using our computed value of $f_B$, we can compute analytically the critical value of the tension that will buckle a ring  in a contractile space, assuming that space to be ellipsoid and near-spherical. For this, we take the very same approach as we did for the bundle in a prolate ellipsoid, although now the ellipsoid is oblate, and the buckling force is that of a ring rather than an open bundle.

\begin{equation}
\sigma^* =\frac{f_B}{ \partial_R S(R,R,R_0^3/R^2) }
\label{critical_tens}
\end{equation}

\bibliography{bibtex_refs}

\begin{thebibliography}{40}%
\makeatletter
\providecommand \@ifxundefined [1]{%
 \@ifx{#1\undefined}
}%
\providecommand \@ifnum [1]{%
 \ifnum #1\expandafter \@firstoftwo
 \else \expandafter \@secondoftwo
 \fi
}%
\providecommand \@ifx [1]{%
 \ifx #1\expandafter \@firstoftwo
 \else \expandafter \@secondoftwo
 \fi
}%
\providecommand \natexlab [1]{#1}%
\providecommand \enquote  [1]{``#1''}%
\providecommand \bibnamefont  [1]{#1}%
\providecommand \bibfnamefont [1]{#1}%
\providecommand \citenamefont [1]{#1}%
\providecommand \href@noop [0]{\@secondoftwo}%
\providecommand \href [0]{\begingroup \@sanitize@url \@href}%
\providecommand \@href[1]{\@@startlink{#1}\@@href}%
\providecommand \@@href[1]{\endgroup#1\@@endlink}%
\providecommand \@sanitize@url [0]{\catcode `\\12\catcode `\$12\catcode
  `\&12\catcode `\#12\catcode `\^12\catcode `\_12\catcode `\%12\relax}%
\providecommand \@@startlink[1]{}%
\providecommand \@@endlink[0]{}%
\providecommand \url  [0]{\begingroup\@sanitize@url \@url }%
\providecommand \@url [1]{\endgroup\@href {#1}{\urlprefix }}%
\providecommand \urlprefix  [0]{URL }%
\providecommand \Eprint [0]{\href }%
\providecommand \doibase [0]{http://dx.doi.org/}%
\providecommand \selectlanguage [0]{\@gobble}%
\providecommand \bibinfo  [0]{\@secondoftwo}%
\providecommand \bibfield  [0]{\@secondoftwo}%
\providecommand \translation [1]{[#1]}%
\providecommand \BibitemOpen [0]{}%
\providecommand \bibitemStop [0]{}%
\providecommand \bibitemNoStop [0]{.\EOS\space}%
\providecommand \EOS [0]{\spacefactor3000\relax}%
\providecommand \BibitemShut  [1]{\csname bibitem#1\endcsname}%
\let\auto@bib@innerbib\@empty
%</preamble>
\bibitem [{\citenamefont {Lecuit}\ and\ \citenamefont
  {Lenne}(2007)}]{lecuit2007cell}%
  \BibitemOpen
  \bibfield  {author} {\bibinfo {author} {\bibfnamefont {T.}~\bibnamefont
  {Lecuit}}\ and\ \bibinfo {author} {\bibfnamefont {P.-F.}\ \bibnamefont
  {Lenne}},\ }\href@noop {} {\bibfield  {journal} {\bibinfo  {journal} {Nature
  Reviews Molecular Cell Biology}\ }\textbf {\bibinfo {volume} {8}},\ \bibinfo
  {pages} {633} (\bibinfo {year} {2007})}\BibitemShut {NoStop}%
\bibitem [{\citenamefont {Goniakowska-Witalinska}\ and\ \citenamefont
  {Witalinski}(1976)}]{goniakowska1976evidence}%
  \BibitemOpen
  \bibfield  {author} {\bibinfo {author} {\bibfnamefont {L.}~\bibnamefont
  {Goniakowska-Witalinska}}\ and\ \bibinfo {author} {\bibfnamefont
  {W.}~\bibnamefont {Witalinski}},\ }\href@noop {} {\bibfield  {journal}
  {\bibinfo  {journal} {Journal of cell science}\ }\textbf {\bibinfo {volume}
  {22}},\ \bibinfo {pages} {397} (\bibinfo {year} {1976})}\BibitemShut
  {NoStop}%
\bibitem [{\citenamefont {Lee}\ \emph {et~al.}(2004)\citenamefont {Lee},
  \citenamefont {Miller}, \citenamefont {Anastassov},\ and\ \citenamefont
  {Cohen}}]{lee2004shape}%
  \BibitemOpen
  \bibfield  {author} {\bibinfo {author} {\bibfnamefont {K.-G.}\ \bibnamefont
  {Lee}}, \bibinfo {author} {\bibfnamefont {T.}~\bibnamefont {Miller}},
  \bibinfo {author} {\bibfnamefont {I.}~\bibnamefont {Anastassov}}, \ and\
  \bibinfo {author} {\bibfnamefont {W.~D.}\ \bibnamefont {Cohen}},\ }\href@noop
  {} {\bibfield  {journal} {\bibinfo  {journal} {Cell biology international}\
  }\textbf {\bibinfo {volume} {28}},\ \bibinfo {pages} {299} (\bibinfo {year}
  {2004})}\BibitemShut {NoStop}%
\bibitem [{\citenamefont {Patel-Hett}\ \emph {et~al.}(2008)\citenamefont
  {Patel-Hett}, \citenamefont {Richardson}, \citenamefont {Schulze},
  \citenamefont {Drabek}, \citenamefont {Isaac}, \citenamefont {Hoffmeister},
  \citenamefont {Shivdasani}, \citenamefont {Bulinski}, \citenamefont
  {Galjart}, \citenamefont {Hartwig} \emph {et~al.}}]{patel2008visualization}%
  \BibitemOpen
  \bibfield  {author} {\bibinfo {author} {\bibfnamefont {S.}~\bibnamefont
  {Patel-Hett}}, \bibinfo {author} {\bibfnamefont {J.~L.}\ \bibnamefont
  {Richardson}}, \bibinfo {author} {\bibfnamefont {H.}~\bibnamefont {Schulze}},
  \bibinfo {author} {\bibfnamefont {K.}~\bibnamefont {Drabek}}, \bibinfo
  {author} {\bibfnamefont {N.~A.}\ \bibnamefont {Isaac}}, \bibinfo {author}
  {\bibfnamefont {K.}~\bibnamefont {Hoffmeister}}, \bibinfo {author}
  {\bibfnamefont {R.~A.}\ \bibnamefont {Shivdasani}}, \bibinfo {author}
  {\bibfnamefont {J.~C.}\ \bibnamefont {Bulinski}}, \bibinfo {author}
  {\bibfnamefont {N.}~\bibnamefont {Galjart}}, \bibinfo {author} {\bibfnamefont
  {J.~H.}\ \bibnamefont {Hartwig}},  \emph {et~al.},\ }\href@noop {} {\bibfield
   {journal} {\bibinfo  {journal} {Blood}\ }\textbf {\bibinfo {volume} {111}},\
  \bibinfo {pages} {4605} (\bibinfo {year} {2008})}\BibitemShut {NoStop}%
\bibitem [{\citenamefont {Van~Deurs}\ and\ \citenamefont
  {Behnke}(1973)}]{van1973microtubule}%
  \BibitemOpen
  \bibfield  {author} {\bibinfo {author} {\bibfnamefont {B.}~\bibnamefont
  {Van~Deurs}}\ and\ \bibinfo {author} {\bibfnamefont {O.}~\bibnamefont
  {Behnke}},\ }\href@noop {} {\bibfield  {journal} {\bibinfo  {journal}
  {Zeitschrift f{\"u}r Anatomie und Entwicklungsgeschichte}\ }\textbf {\bibinfo
  {volume} {143}},\ \bibinfo {pages} {43} (\bibinfo {year} {1973})}\BibitemShut
  {NoStop}%
\bibitem [{\citenamefont {White}(2013)}]{white2013}%
  \BibitemOpen
  \bibfield  {author} {\bibinfo {author} {\bibfnamefont {J.~G.}\ \bibnamefont
  {White}},\ }in\ \href@noop {} {\emph {\bibinfo {booktitle} {Platelets, third
  edition}}},\ \bibinfo {editor} {edited by\ \bibinfo {editor} {\bibfnamefont
  {A.~D.}\ \bibnamefont {Michelson}}}\ (\bibinfo  {publisher} {Academic
  Press},\ \bibinfo {year} {2013})\ Chap.~\bibinfo {chapter} {7}, pp.\ \bibinfo
  {pages} {117--144}\BibitemShut {NoStop}%
\bibitem [{\citenamefont {Joseph-Silverstein}\ and\ \citenamefont
  {Cohen}(1984)}]{joseph1984cytoskeletal}%
  \BibitemOpen
  \bibfield  {author} {\bibinfo {author} {\bibfnamefont {J.}~\bibnamefont
  {Joseph-Silverstein}}\ and\ \bibinfo {author} {\bibfnamefont {W.~D.}\
  \bibnamefont {Cohen}},\ }\href@noop {} {\bibfield  {journal} {\bibinfo
  {journal} {The Journal of cell biology}\ }\textbf {\bibinfo {volume} {98}},\
  \bibinfo {pages} {2118} (\bibinfo {year} {1984})}\BibitemShut {NoStop}%
\bibitem [{\citenamefont {Schroter}\ \emph {et~al.}(1990)\citenamefont
  {Schroter}, \citenamefont {Filali}, \citenamefont {Brain}, \citenamefont
  {Jeffrey},\ and\ \citenamefont {Robertshaw}}]{schroter1990influence}%
  \BibitemOpen
  \bibfield  {author} {\bibinfo {author} {\bibfnamefont {R.}~\bibnamefont
  {Schroter}}, \bibinfo {author} {\bibfnamefont {R.~Z.}\ \bibnamefont
  {Filali}}, \bibinfo {author} {\bibfnamefont {A.}~\bibnamefont {Brain}},
  \bibinfo {author} {\bibfnamefont {P.}~\bibnamefont {Jeffrey}}, \ and\
  \bibinfo {author} {\bibfnamefont {D.}~\bibnamefont {Robertshaw}},\
  }\href@noop {} {\bibfield  {journal} {\bibinfo  {journal} {Respiration
  physiology}\ }\textbf {\bibinfo {volume} {81}},\ \bibinfo {pages} {381}
  (\bibinfo {year} {1990})}\BibitemShut {NoStop}%
\bibitem [{\citenamefont {Hartwig}\ and\ \citenamefont
  {DeSisto}(1991)}]{hartwig1991cytoskeleton}%
  \BibitemOpen
  \bibfield  {author} {\bibinfo {author} {\bibfnamefont {J.~H.}\ \bibnamefont
  {Hartwig}}\ and\ \bibinfo {author} {\bibfnamefont {M.}~\bibnamefont
  {DeSisto}},\ }\href@noop {} {\bibfield  {journal} {\bibinfo  {journal} {The
  Journal of cell biology}\ }\textbf {\bibinfo {volume} {112}},\ \bibinfo
  {pages} {407} (\bibinfo {year} {1991})}\BibitemShut {NoStop}%
\bibitem [{\citenamefont {Bathe}\ \emph {et~al.}(2008)\citenamefont {Bathe},
  \citenamefont {Heussinger}, \citenamefont {Claessens}, \citenamefont
  {Bausch},\ and\ \citenamefont {Frey}}]{bathe2008cytoskeletal}%
  \BibitemOpen
  \bibfield  {author} {\bibinfo {author} {\bibfnamefont {M.}~\bibnamefont
  {Bathe}}, \bibinfo {author} {\bibfnamefont {C.}~\bibnamefont {Heussinger}},
  \bibinfo {author} {\bibfnamefont {M.~M.}\ \bibnamefont {Claessens}}, \bibinfo
  {author} {\bibfnamefont {A.~R.}\ \bibnamefont {Bausch}}, \ and\ \bibinfo
  {author} {\bibfnamefont {E.}~\bibnamefont {Frey}},\ }\href@noop {} {\bibfield
   {journal} {\bibinfo  {journal} {Biophysical journal}\ }\textbf {\bibinfo
  {volume} {94}},\ \bibinfo {pages} {2955} (\bibinfo {year}
  {2008})}\BibitemShut {NoStop}%
\bibitem [{\citenamefont {Patel-Hett}\ \emph {et~al.}(2011)\citenamefont
  {Patel-Hett}, \citenamefont {Wang}, \citenamefont {Begonja}, \citenamefont
  {Thon}, \citenamefont {Alden}, \citenamefont {Wandersee}, \citenamefont {An},
  \citenamefont {Mohandas}, \citenamefont {Hartwig},\ and\ \citenamefont
  {Italiano}}]{patel2011spectrin}%
  \BibitemOpen
  \bibfield  {author} {\bibinfo {author} {\bibfnamefont {S.}~\bibnamefont
  {Patel-Hett}}, \bibinfo {author} {\bibfnamefont {H.}~\bibnamefont {Wang}},
  \bibinfo {author} {\bibfnamefont {A.~J.}\ \bibnamefont {Begonja}}, \bibinfo
  {author} {\bibfnamefont {J.~N.}\ \bibnamefont {Thon}}, \bibinfo {author}
  {\bibfnamefont {E.~C.}\ \bibnamefont {Alden}}, \bibinfo {author}
  {\bibfnamefont {N.~J.}\ \bibnamefont {Wandersee}}, \bibinfo {author}
  {\bibfnamefont {X.}~\bibnamefont {An}}, \bibinfo {author} {\bibfnamefont
  {N.}~\bibnamefont {Mohandas}}, \bibinfo {author} {\bibfnamefont {J.~H.}\
  \bibnamefont {Hartwig}}, \ and\ \bibinfo {author} {\bibfnamefont {J.~E.}\
  \bibnamefont {Italiano}},\ }\href@noop {} {\bibfield  {journal} {\bibinfo
  {journal} {Blood}\ }\textbf {\bibinfo {volume} {118}},\ \bibinfo {pages}
  {1641} (\bibinfo {year} {2011})}\BibitemShut {NoStop}%
\bibitem [{\citenamefont {Thon}\ \emph {et~al.}(2012)\citenamefont {Thon},
  \citenamefont {Macleod}, \citenamefont {Begonja}, \citenamefont {Zhu},
  \citenamefont {Lee}, \citenamefont {Mogilner}, \citenamefont {Hartwig},\ and\
  \citenamefont {Italiano~Jr}}]{thon2012microtubule}%
  \BibitemOpen
  \bibfield  {author} {\bibinfo {author} {\bibfnamefont {J.~N.}\ \bibnamefont
  {Thon}}, \bibinfo {author} {\bibfnamefont {H.}~\bibnamefont {Macleod}},
  \bibinfo {author} {\bibfnamefont {A.~J.}\ \bibnamefont {Begonja}}, \bibinfo
  {author} {\bibfnamefont {J.}~\bibnamefont {Zhu}}, \bibinfo {author}
  {\bibfnamefont {K.-C.}\ \bibnamefont {Lee}}, \bibinfo {author} {\bibfnamefont
  {A.}~\bibnamefont {Mogilner}}, \bibinfo {author} {\bibfnamefont {J.~H.}\
  \bibnamefont {Hartwig}}, \ and\ \bibinfo {author} {\bibfnamefont {J.~E.}\
  \bibnamefont {Italiano~Jr}},\ }\href@noop {} {\bibfield  {journal} {\bibinfo
  {journal} {Nature communications}\ }\textbf {\bibinfo {volume} {3}},\
  \bibinfo {pages} {852} (\bibinfo {year} {2012})}\BibitemShut {NoStop}%
\bibitem [{\citenamefont {Cohen}\ \emph {et~al.}(1982)\citenamefont {Cohen},
  \citenamefont {Bartelt}, \citenamefont {Jaeger}, \citenamefont {Langford},\
  and\ \citenamefont {Nemhauser}}]{cohen1982cytoskeletal}%
  \BibitemOpen
  \bibfield  {author} {\bibinfo {author} {\bibfnamefont {W.~D.}\ \bibnamefont
  {Cohen}}, \bibinfo {author} {\bibfnamefont {D.}~\bibnamefont {Bartelt}},
  \bibinfo {author} {\bibfnamefont {R.}~\bibnamefont {Jaeger}}, \bibinfo
  {author} {\bibfnamefont {G.}~\bibnamefont {Langford}}, \ and\ \bibinfo
  {author} {\bibfnamefont {I.}~\bibnamefont {Nemhauser}},\ }\href@noop {}
  {\bibfield  {journal} {\bibinfo  {journal} {The Journal of cell biology}\
  }\textbf {\bibinfo {volume} {93}},\ \bibinfo {pages} {828} (\bibinfo {year}
  {1982})}\BibitemShut {NoStop}%
\bibitem [{\citenamefont {Evans}\ and\ \citenamefont
  {Yeung}(1989)}]{evans1989apparent}%
  \BibitemOpen
  \bibfield  {author} {\bibinfo {author} {\bibfnamefont {E.}~\bibnamefont
  {Evans}}\ and\ \bibinfo {author} {\bibfnamefont {A.}~\bibnamefont {Yeung}},\
  }\href@noop {} {\bibfield  {journal} {\bibinfo  {journal} {Biophysical
  journal}\ }\textbf {\bibinfo {volume} {56}},\ \bibinfo {pages} {151}
  (\bibinfo {year} {1989})}\BibitemShut {NoStop}%
\bibitem [{\citenamefont {Stewart}\ \emph {et~al.}(2011)\citenamefont
  {Stewart}, \citenamefont {Helenius}, \citenamefont {Toyoda}, \citenamefont
  {Ramanathan}, \citenamefont {Muller},\ and\ \citenamefont
  {Hyman}}]{stewart2011hydrostatic}%
  \BibitemOpen
  \bibfield  {author} {\bibinfo {author} {\bibfnamefont {M.~P.}\ \bibnamefont
  {Stewart}}, \bibinfo {author} {\bibfnamefont {J.}~\bibnamefont {Helenius}},
  \bibinfo {author} {\bibfnamefont {Y.}~\bibnamefont {Toyoda}}, \bibinfo
  {author} {\bibfnamefont {S.~P.}\ \bibnamefont {Ramanathan}}, \bibinfo
  {author} {\bibfnamefont {D.~J.}\ \bibnamefont {Muller}}, \ and\ \bibinfo
  {author} {\bibfnamefont {A.~A.}\ \bibnamefont {Hyman}},\ }\href@noop {}
  {\bibfield  {journal} {\bibinfo  {journal} {Nature}\ }\textbf {\bibinfo
  {volume} {469}},\ \bibinfo {pages} {226} (\bibinfo {year}
  {2011})}\BibitemShut {NoStop}%
\bibitem [{\citenamefont {Bender}\ \emph {et~al.}(2015)\citenamefont {Bender},
  \citenamefont {Thon}, \citenamefont {Ehrlicher}, \citenamefont {Wu},
  \citenamefont {Mazutis}, \citenamefont {Deschmann}, \citenamefont
  {Sola-Visner}, \citenamefont {Italiano},\ and\ \citenamefont
  {Hartwig}}]{bender2015microtubule}%
  \BibitemOpen
  \bibfield  {author} {\bibinfo {author} {\bibfnamefont {M.}~\bibnamefont
  {Bender}}, \bibinfo {author} {\bibfnamefont {J.~N.}\ \bibnamefont {Thon}},
  \bibinfo {author} {\bibfnamefont {A.~J.}\ \bibnamefont {Ehrlicher}}, \bibinfo
  {author} {\bibfnamefont {S.}~\bibnamefont {Wu}}, \bibinfo {author}
  {\bibfnamefont {L.}~\bibnamefont {Mazutis}}, \bibinfo {author} {\bibfnamefont
  {E.}~\bibnamefont {Deschmann}}, \bibinfo {author} {\bibfnamefont
  {M.}~\bibnamefont {Sola-Visner}}, \bibinfo {author} {\bibfnamefont {J.~E.}\
  \bibnamefont {Italiano}}, \ and\ \bibinfo {author} {\bibfnamefont {J.~H.}\
  \bibnamefont {Hartwig}},\ }\href@noop {} {\bibfield  {journal} {\bibinfo
  {journal} {Blood}\ }\textbf {\bibinfo {volume} {125}},\ \bibinfo {pages}
  {860} (\bibinfo {year} {2015})}\BibitemShut {NoStop}%
\bibitem [{\citenamefont {White}\ and\ \citenamefont
  {Rao}(1998)}]{white1998microtubule}%
  \BibitemOpen
  \bibfield  {author} {\bibinfo {author} {\bibfnamefont {J.~G.}\ \bibnamefont
  {White}}\ and\ \bibinfo {author} {\bibfnamefont {G.}~\bibnamefont {Rao}},\
  }\href@noop {} {\bibfield  {journal} {\bibinfo  {journal} {The American
  journal of pathology}\ }\textbf {\bibinfo {volume} {152}},\ \bibinfo {pages}
  {597} (\bibinfo {year} {1998})}\BibitemShut {NoStop}%
\bibitem [{\citenamefont {Ostermeir}\ \emph {et~al.}(2010)\citenamefont
  {Ostermeir}, \citenamefont {Alim},\ and\ \citenamefont
  {Frey}}]{ostermeir2010buckling}%
  \BibitemOpen
  \bibfield  {author} {\bibinfo {author} {\bibfnamefont {K.}~\bibnamefont
  {Ostermeir}}, \bibinfo {author} {\bibfnamefont {K.}~\bibnamefont {Alim}}, \
  and\ \bibinfo {author} {\bibfnamefont {E.}~\bibnamefont {Frey}},\ }\href@noop
  {} {\bibfield  {journal} {\bibinfo  {journal} {Physical Review E}\ }\textbf
  {\bibinfo {volume} {81}},\ \bibinfo {pages} {061802} (\bibinfo {year}
  {2010})}\BibitemShut {NoStop}%
\bibitem [{\citenamefont {Diagouraga}\ \emph {et~al.}(2014)\citenamefont
  {Diagouraga}, \citenamefont {Grichine}, \citenamefont {Fertin}, \citenamefont
  {Wang}, \citenamefont {Khochbin},\ and\ \citenamefont
  {Sadoul}}]{diagouraga2014motor}%
  \BibitemOpen
  \bibfield  {author} {\bibinfo {author} {\bibfnamefont {B.}~\bibnamefont
  {Diagouraga}}, \bibinfo {author} {\bibfnamefont {A.}~\bibnamefont
  {Grichine}}, \bibinfo {author} {\bibfnamefont {A.}~\bibnamefont {Fertin}},
  \bibinfo {author} {\bibfnamefont {J.}~\bibnamefont {Wang}}, \bibinfo {author}
  {\bibfnamefont {S.}~\bibnamefont {Khochbin}}, \ and\ \bibinfo {author}
  {\bibfnamefont {K.}~\bibnamefont {Sadoul}},\ }\href@noop {} {\bibfield
  {journal} {\bibinfo  {journal} {The Journal of cell biology}\ }\textbf
  {\bibinfo {volume} {204}},\ \bibinfo {pages} {177} (\bibinfo {year}
  {2014})}\BibitemShut {NoStop}%
\bibitem [{\citenamefont {Kuwahara}\ \emph {et~al.}(2002)\citenamefont
  {Kuwahara}, \citenamefont {Sugimoto}, \citenamefont {Tsuji}, \citenamefont
  {Matsui}, \citenamefont {Mizuno}, \citenamefont {Miyata},\ and\ \citenamefont
  {Yoshioka}}]{kuwahara2002platelet}%
  \BibitemOpen
  \bibfield  {author} {\bibinfo {author} {\bibfnamefont {M.}~\bibnamefont
  {Kuwahara}}, \bibinfo {author} {\bibfnamefont {M.}~\bibnamefont {Sugimoto}},
  \bibinfo {author} {\bibfnamefont {S.}~\bibnamefont {Tsuji}}, \bibinfo
  {author} {\bibfnamefont {H.}~\bibnamefont {Matsui}}, \bibinfo {author}
  {\bibfnamefont {T.}~\bibnamefont {Mizuno}}, \bibinfo {author} {\bibfnamefont
  {S.}~\bibnamefont {Miyata}}, \ and\ \bibinfo {author} {\bibfnamefont
  {A.}~\bibnamefont {Yoshioka}},\ }\href@noop {} {\bibfield  {journal}
  {\bibinfo  {journal} {Arteriosclerosis, thrombosis, and vascular biology}\
  }\textbf {\bibinfo {volume} {22}},\ \bibinfo {pages} {329} (\bibinfo {year}
  {2002})}\BibitemShut {NoStop}%
\bibitem [{\citenamefont {Braun}\ \emph {et~al.}(2011)\citenamefont {Braun},
  \citenamefont {Lansky}, \citenamefont {Fink}, \citenamefont {Ruhnow},
  \citenamefont {Diez},\ and\ \citenamefont {Janson}}]{braun2011adaptive}%
  \BibitemOpen
  \bibfield  {author} {\bibinfo {author} {\bibfnamefont {M.}~\bibnamefont
  {Braun}}, \bibinfo {author} {\bibfnamefont {Z.}~\bibnamefont {Lansky}},
  \bibinfo {author} {\bibfnamefont {G.}~\bibnamefont {Fink}}, \bibinfo {author}
  {\bibfnamefont {F.}~\bibnamefont {Ruhnow}}, \bibinfo {author} {\bibfnamefont
  {S.}~\bibnamefont {Diez}}, \ and\ \bibinfo {author} {\bibfnamefont {M.~E.}\
  \bibnamefont {Janson}},\ }\href@noop {} {\bibfield  {journal} {\bibinfo
  {journal} {Nature cell biology}\ }\textbf {\bibinfo {volume} {13}},\ \bibinfo
  {pages} {1259} (\bibinfo {year} {2011})}\BibitemShut {NoStop}%
\bibitem [{\citenamefont {Gittes}\ \emph {et~al.}(1993)\citenamefont {Gittes},
  \citenamefont {Mickey}, \citenamefont {Nettleton},\ and\ \citenamefont
  {Howard}}]{gittes1993flexural}%
  \BibitemOpen
  \bibfield  {author} {\bibinfo {author} {\bibfnamefont {F.}~\bibnamefont
  {Gittes}}, \bibinfo {author} {\bibfnamefont {B.}~\bibnamefont {Mickey}},
  \bibinfo {author} {\bibfnamefont {J.}~\bibnamefont {Nettleton}}, \ and\
  \bibinfo {author} {\bibfnamefont {J.}~\bibnamefont {Howard}},\ }\href@noop {}
  {\bibfield  {journal} {\bibinfo  {journal} {The Journal of cell biology}\
  }\textbf {\bibinfo {volume} {120}},\ \bibinfo {pages} {923} (\bibinfo {year}
  {1993})}\BibitemShut {NoStop}%
\bibitem [{\citenamefont {Salbreux}\ \emph {et~al.}(2012)\citenamefont
  {Salbreux}, \citenamefont {Charras},\ and\ \citenamefont
  {Paluch}}]{salbreux2012actin}%
  \BibitemOpen
  \bibfield  {author} {\bibinfo {author} {\bibfnamefont {G.}~\bibnamefont
  {Salbreux}}, \bibinfo {author} {\bibfnamefont {G.}~\bibnamefont {Charras}}, \
  and\ \bibinfo {author} {\bibfnamefont {E.}~\bibnamefont {Paluch}},\
  }\href@noop {} {\bibfield  {journal} {\bibinfo  {journal} {Trends in cell
  biology}\ }\textbf {\bibinfo {volume} {22}},\ \bibinfo {pages} {536}
  (\bibinfo {year} {2012})}\BibitemShut {NoStop}%
\bibitem [{\citenamefont {Tinevez}\ \emph {et~al.}(2009)\citenamefont
  {Tinevez}, \citenamefont {Schulze}, \citenamefont {Salbreux}, \citenamefont
  {Roensch}, \citenamefont {Joanny},\ and\ \citenamefont
  {Paluch}}]{tinevez2009role}%
  \BibitemOpen
  \bibfield  {author} {\bibinfo {author} {\bibfnamefont {J.-Y.}\ \bibnamefont
  {Tinevez}}, \bibinfo {author} {\bibfnamefont {U.}~\bibnamefont {Schulze}},
  \bibinfo {author} {\bibfnamefont {G.}~\bibnamefont {Salbreux}}, \bibinfo
  {author} {\bibfnamefont {J.}~\bibnamefont {Roensch}}, \bibinfo {author}
  {\bibfnamefont {J.-F.}\ \bibnamefont {Joanny}}, \ and\ \bibinfo {author}
  {\bibfnamefont {E.}~\bibnamefont {Paluch}},\ }\href@noop {} {\bibfield
  {journal} {\bibinfo  {journal} {Proceedings of the National Academy of
  Sciences}\ }\textbf {\bibinfo {volume} {106}},\ \bibinfo {pages} {18581}
  (\bibinfo {year} {2009})}\BibitemShut {NoStop}%
\bibitem [{\citenamefont {Fournier}\ \emph {et~al.}(2004)\citenamefont
  {Fournier}, \citenamefont {Lacoste},\ and\ \citenamefont
  {Rapha{\"e}l}}]{fournier2004fluctuation}%
  \BibitemOpen
  \bibfield  {author} {\bibinfo {author} {\bibfnamefont {J.-B.}\ \bibnamefont
  {Fournier}}, \bibinfo {author} {\bibfnamefont {D.}~\bibnamefont {Lacoste}}, \
  and\ \bibinfo {author} {\bibfnamefont {E.}~\bibnamefont {Rapha{\"e}l}},\
  }\href@noop {} {\bibfield  {journal} {\bibinfo  {journal} {Physical review
  letters}\ }\textbf {\bibinfo {volume} {92}},\ \bibinfo {pages} {018102}
  (\bibinfo {year} {2004})}\BibitemShut {NoStop}%
\bibitem [{\citenamefont {Turlier}\ \emph {et~al.}(2016)\citenamefont
  {Turlier}, \citenamefont {Fedosov}, \citenamefont {Audoly}, \citenamefont
  {Auth}, \citenamefont {Gov}, \citenamefont {Sykes}, \citenamefont {Joanny},
  \citenamefont {Gompper},\ and\ \citenamefont
  {Betz}}]{turlier2016equilibrium}%
  \BibitemOpen
  \bibfield  {author} {\bibinfo {author} {\bibfnamefont {H.}~\bibnamefont
  {Turlier}}, \bibinfo {author} {\bibfnamefont {D.}~\bibnamefont {Fedosov}},
  \bibinfo {author} {\bibfnamefont {B.}~\bibnamefont {Audoly}}, \bibinfo
  {author} {\bibfnamefont {T.}~\bibnamefont {Auth}}, \bibinfo {author}
  {\bibfnamefont {N.}~\bibnamefont {Gov}}, \bibinfo {author} {\bibfnamefont
  {C.}~\bibnamefont {Sykes}}, \bibinfo {author} {\bibfnamefont {J.-F.}\
  \bibnamefont {Joanny}}, \bibinfo {author} {\bibfnamefont {G.}~\bibnamefont
  {Gompper}}, \ and\ \bibinfo {author} {\bibfnamefont {T.}~\bibnamefont
  {Betz}},\ }\href@noop {} {\bibfield  {journal} {\bibinfo  {journal} {Nature
  Physics}\ } (\bibinfo {year} {2016})}\BibitemShut {NoStop}%
\bibitem [{\citenamefont {Kenney}\ and\ \citenamefont
  {Linck}(1985)}]{kenney1985cystoskeleton}%
  \BibitemOpen
  \bibfield  {author} {\bibinfo {author} {\bibfnamefont {D.~M.}\ \bibnamefont
  {Kenney}}\ and\ \bibinfo {author} {\bibfnamefont {R.}~\bibnamefont {Linck}},\
  }\href@noop {} {\bibfield  {journal} {\bibinfo  {journal} {Journal of cell
  science}\ }\textbf {\bibinfo {volume} {78}},\ \bibinfo {pages} {1} (\bibinfo
  {year} {1985})}\BibitemShut {NoStop}%
\bibitem [{\citenamefont {Nedelec}\ and\ \citenamefont
  {Foethke}(2007)}]{nedelec2007collective}%
  \BibitemOpen
  \bibfield  {author} {\bibinfo {author} {\bibfnamefont {F.}~\bibnamefont
  {Nedelec}}\ and\ \bibinfo {author} {\bibfnamefont {D.}~\bibnamefont
  {Foethke}},\ }\href@noop {} {\bibfield  {journal} {\bibinfo  {journal} {New
  Journal of Physics}\ }\textbf {\bibinfo {volume} {9}},\ \bibinfo {pages}
  {427} (\bibinfo {year} {2007})}\BibitemShut {NoStop}%
\bibitem [{\citenamefont {Hartwig}(2002)}]{hartwig2002platelet}%
  \BibitemOpen
  \bibfield  {author} {\bibinfo {author} {\bibfnamefont {J.~H.}\ \bibnamefont
  {Hartwig}},\ }\href@noop {} {\bibfield  {journal} {\bibinfo  {journal}
  {Platelets}\ }\textbf {\bibinfo {volume} {1}},\ \bibinfo {pages} {26}
  (\bibinfo {year} {2002})}\BibitemShut {NoStop}%
\bibitem [{\citenamefont {Cohen}(1991)}]{cohen1991cytoskeletal}%
  \BibitemOpen
  \bibfield  {author} {\bibinfo {author} {\bibfnamefont {W.~D.}\ \bibnamefont
  {Cohen}},\ }\href@noop {} {\bibfield  {journal} {\bibinfo  {journal} {Int Rev
  Cytol}\ }\textbf {\bibinfo {volume} {130}},\ \bibinfo {pages} {37} (\bibinfo
  {year} {1991})}\BibitemShut {NoStop}%
\bibitem [{\citenamefont {Guven}\ and\ \citenamefont
  {V{\'a}zquez-Montejo}(2012)}]{guven2012confinement}%
  \BibitemOpen
  \bibfield  {author} {\bibinfo {author} {\bibfnamefont {J.}~\bibnamefont
  {Guven}}\ and\ \bibinfo {author} {\bibfnamefont {P.}~\bibnamefont
  {V{\'a}zquez-Montejo}},\ }\href@noop {} {\bibfield  {journal} {\bibinfo
  {journal} {Physical Review E}\ }\textbf {\bibinfo {volume} {85}},\ \bibinfo
  {pages} {026603} (\bibinfo {year} {2012})}\BibitemShut {NoStop}%
\bibitem [{\citenamefont {Lux}\ \emph {et~al.}(1978)\citenamefont {Lux},
  \citenamefont {John},\ and\ \citenamefont {Ukena}}]{lux1978diminished}%
  \BibitemOpen
  \bibfield  {author} {\bibinfo {author} {\bibfnamefont {S.}~\bibnamefont
  {Lux}}, \bibinfo {author} {\bibfnamefont {K.}~\bibnamefont {John}}, \ and\
  \bibinfo {author} {\bibfnamefont {T.~E.}\ \bibnamefont {Ukena}},\ }\href@noop
  {} {\bibfield  {journal} {\bibinfo  {journal} {Journal of Clinical
  Investigation}\ }\textbf {\bibinfo {volume} {61}},\ \bibinfo {pages} {815}
  (\bibinfo {year} {1978})}\BibitemShut {NoStop}%
\bibitem [{\citenamefont {Cohen}\ \emph {et~al.}(1998)\citenamefont {Cohen},
  \citenamefont {Sorokina},\ and\ \citenamefont
  {Sanchez}}]{cohen1998elliptical}%
  \BibitemOpen
  \bibfield  {author} {\bibinfo {author} {\bibfnamefont {W.~D.}\ \bibnamefont
  {Cohen}}, \bibinfo {author} {\bibfnamefont {Y.}~\bibnamefont {Sorokina}}, \
  and\ \bibinfo {author} {\bibfnamefont {I.}~\bibnamefont {Sanchez}},\
  }\href@noop {} {\bibfield  {journal} {\bibinfo  {journal} {Cell motility and
  the cytoskeleton}\ }\textbf {\bibinfo {volume} {40}},\ \bibinfo {pages} {238}
  (\bibinfo {year} {1998})}\BibitemShut {NoStop}%
\bibitem [{\citenamefont {Ward}\ \emph {et~al.}(2014)\citenamefont {Ward},
  \citenamefont {Roque}, \citenamefont {Antony},\ and\ \citenamefont
  {N{\'e}d{\'e}lec}}]{ward2014mechanical}%
  \BibitemOpen
  \bibfield  {author} {\bibinfo {author} {\bibfnamefont {J.~J.}\ \bibnamefont
  {Ward}}, \bibinfo {author} {\bibfnamefont {H.}~\bibnamefont {Roque}},
  \bibinfo {author} {\bibfnamefont {C.}~\bibnamefont {Antony}}, \ and\ \bibinfo
  {author} {\bibfnamefont {F.}~\bibnamefont {N{\'e}d{\'e}lec}},\ }\href@noop {}
  {\bibfield  {journal} {\bibinfo  {journal} {Elife}\ }\textbf {\bibinfo
  {volume} {3}},\ \bibinfo {pages} {e03398} (\bibinfo {year}
  {2014})}\BibitemShut {NoStop}%
\bibitem [{\citenamefont {Hartwig}(1992)}]{hartwig1992mechanisms}%
  \BibitemOpen
  \bibfield  {author} {\bibinfo {author} {\bibfnamefont {J.~H.}\ \bibnamefont
  {Hartwig}},\ }\href@noop {} {\bibfield  {journal} {\bibinfo  {journal} {The
  Journal of Cell Biology}\ }\textbf {\bibinfo {volume} {118}},\ \bibinfo
  {pages} {1421} (\bibinfo {year} {1992})}\BibitemShut {NoStop}%
\bibitem [{\citenamefont {Carroll}\ and\ \citenamefont
  {Gerrard}(1982)}]{carroll1982phosphorylation}%
  \BibitemOpen
  \bibfield  {author} {\bibinfo {author} {\bibfnamefont {R.~C.}\ \bibnamefont
  {Carroll}}\ and\ \bibinfo {author} {\bibfnamefont {J.~M.}\ \bibnamefont
  {Gerrard}},\ }\href@noop {} {\bibfield  {journal} {\bibinfo  {journal}
  {Blood}\ }\textbf {\bibinfo {volume} {59}},\ \bibinfo {pages} {466} (\bibinfo
  {year} {1982})}\BibitemShut {NoStop}%
\bibitem [{\citenamefont {Li}\ \emph {et~al.}(2002)\citenamefont {Li},
  \citenamefont {Kim},\ and\ \citenamefont {Bearer}}]{li2002arp2}%
  \BibitemOpen
  \bibfield  {author} {\bibinfo {author} {\bibfnamefont {Z.}~\bibnamefont
  {Li}}, \bibinfo {author} {\bibfnamefont {E.~S.}\ \bibnamefont {Kim}}, \ and\
  \bibinfo {author} {\bibfnamefont {E.~L.}\ \bibnamefont {Bearer}},\
  }\href@noop {} {\bibfield  {journal} {\bibinfo  {journal} {Blood}\ }\textbf
  {\bibinfo {volume} {99}},\ \bibinfo {pages} {4466} (\bibinfo {year}
  {2002})}\BibitemShut {NoStop}%
\bibitem [{\citenamefont {Lee}\ \emph {et~al.}(2007)\citenamefont {Lee},
  \citenamefont {Johner},\ and\ \citenamefont {Hong}}]{lee2007compressing}%
  \BibitemOpen
  \bibfield  {author} {\bibinfo {author} {\bibfnamefont {N.-K.}\ \bibnamefont
  {Lee}}, \bibinfo {author} {\bibfnamefont {A.}~\bibnamefont {Johner}}, \ and\
  \bibinfo {author} {\bibfnamefont {S.-C.}\ \bibnamefont {Hong}},\ }\href@noop
  {} {\bibfield  {journal} {\bibinfo  {journal} {The European Physical Journal
  E}\ }\textbf {\bibinfo {volume} {24}},\ \bibinfo {pages} {229} (\bibinfo
  {year} {2007})}\BibitemShut {NoStop}%
\bibitem [{\citenamefont {Cazenave}\ \emph {et~al.}(2004)\citenamefont
  {Cazenave}, \citenamefont {Ohlmann}, \citenamefont {Cassel}, \citenamefont
  {Eckly}, \citenamefont {Hechler},\ and\ \citenamefont
  {Gachet}}]{cazenave2004preparation}%
  \BibitemOpen
  \bibfield  {author} {\bibinfo {author} {\bibfnamefont {J.-P.}\ \bibnamefont
  {Cazenave}}, \bibinfo {author} {\bibfnamefont {P.}~\bibnamefont {Ohlmann}},
  \bibinfo {author} {\bibfnamefont {D.}~\bibnamefont {Cassel}}, \bibinfo
  {author} {\bibfnamefont {A.}~\bibnamefont {Eckly}}, \bibinfo {author}
  {\bibfnamefont {B.}~\bibnamefont {Hechler}}, \ and\ \bibinfo {author}
  {\bibfnamefont {C.}~\bibnamefont {Gachet}},\ }\href@noop {} {\bibfield
  {journal} {\bibinfo  {journal} {Platelets and Megakaryocytes: Volume 1:
  Functional Assays}\ ,\ \bibinfo {pages} {13}} (\bibinfo {year}
  {2004})}\BibitemShut {NoStop}%
\bibitem [{\citenamefont {Lukinavi{\v{c}}ius}\ \emph
  {et~al.}(2014)\citenamefont {Lukinavi{\v{c}}ius}, \citenamefont {Reymond},
  \citenamefont {D'Este}, \citenamefont {Masharina}, \citenamefont
  {G{\"o}ttfert}, \citenamefont {Ta}, \citenamefont {G{\"u}ther}, \citenamefont
  {Fournier}, \citenamefont {Rizzo}, \citenamefont {Waldmann} \emph
  {et~al.}}]{lukinavivcius2014fluorogenic}%
  \BibitemOpen
  \bibfield  {author} {\bibinfo {author} {\bibfnamefont {G.}~\bibnamefont
  {Lukinavi{\v{c}}ius}}, \bibinfo {author} {\bibfnamefont {L.}~\bibnamefont
  {Reymond}}, \bibinfo {author} {\bibfnamefont {E.}~\bibnamefont {D'Este}},
  \bibinfo {author} {\bibfnamefont {A.}~\bibnamefont {Masharina}}, \bibinfo
  {author} {\bibfnamefont {F.}~\bibnamefont {G{\"o}ttfert}}, \bibinfo {author}
  {\bibfnamefont {H.}~\bibnamefont {Ta}}, \bibinfo {author} {\bibfnamefont
  {A.}~\bibnamefont {G{\"u}ther}}, \bibinfo {author} {\bibfnamefont
  {M.}~\bibnamefont {Fournier}}, \bibinfo {author} {\bibfnamefont
  {S.}~\bibnamefont {Rizzo}}, \bibinfo {author} {\bibfnamefont
  {H.}~\bibnamefont {Waldmann}},  \emph {et~al.},\ }\href@noop {} {\bibfield
  {journal} {\bibinfo  {journal} {Nature methods}\ }\textbf {\bibinfo {volume}
  {11}},\ \bibinfo {pages} {731} (\bibinfo {year} {2014})}\BibitemShut
  {NoStop}%
\end{thebibliography}%

\end{document}